\documentclass[useAMS,usenatbib]{mn2e}
\usepackage{graphicx}

\title[A dynamical model for the extra-planar gas in spiral
galaxies]{A dynamical model for the extra-planar gas in spiral galaxies}

\author[F. Fraternali and J. J. Binney]
{F. Fraternali\thanks{E-mail:
filippo@thphys.ox.ac.uk} and J. J. Binney\\
Theoretical Physics, 1 Keble Road, Oxford, OX1 3NP, UK\\}

\newcommand {\hi} {H\,{\small I}}
\newcommand {\hii} {H\,{\small II}}
\newcommand {\ha} {H$\alpha$}
\newcommand {\kms} {\,{\rm km\,s}^{-1}}
\newcommand {\kpc} {\,{\rm kpc}}
\newcommand {\de}{^{\circ}}
\newcommand {\mo}{\,{M}_\odot}
\newcommand{\yr}{\,{\rm yr}}
\newcommand {\moyr}{\,{M_\odot\,\rm yr}^{-1}}
\newcommand {\mopc}{M_\odot\,{\rm pc^{-2}}}
\newcommand {\lo}{L_{\odot}}
\newcommand {\loB}{L_{\odot \rm, B}}
\newcommand{\gsim}{\lower.7ex\hbox{$\;\stackrel{\textstyle>}{\sim}\;$}}
\newcommand{\lsim}{\lower.7ex\hbox{$\;\stackrel{\textstyle<}{\sim}\;$}}
\newcommand{\ergs}{\,{\rm erg\,s}^{-1}}
\newcommand {\e}{{\rm e}}
\begin{document}

\date{Accepted xxx. Received xxx}

\pagerange{\pageref{firstpage}--\pageref{lastpage}} \pubyear{xxxx}

\maketitle

\label{firstpage}

\begin{abstract}

Recent \hi\ observations reveal that the discs of spiral galaxies are
surrounded by extended gaseous haloes.
This {\it extra-planar} gas reaches large distances (several kpc) from 
the disc and shows peculiar kinematics (low rotation and inflow). 
We have modelled the extra-planar gas as a continuous flow of material
from the disc of a spiral galaxy into its halo region.
The output of our models are pseudo-data cubes that can be directly
compared to the \hi\ data.
We have applied these models to two spiral galaxies (NGC\,891 and
NGC\,2403) known to have a substantial amount of extra-planar gas.
Our models are able to reproduce accurately the vertical distribution
of extra-planar gas for an energy input corresponding to a small 
fraction ($<$4 \%) of the energy released by supernovae. 
However they fail in two important aspects: 
1) they do not reproduce the right gradient in rotation velocity; 
2) they predict a general outflow of the extra-planar gas, contrary to
   what is observed. 
We show that neither of these difficulties can be removed if
clouds are ionized and invisible at 21$\,$cm as they leave the disc but
become visible at some point on their orbits. 
We speculate that these failures indicate the need for accreted
material from the IGM that could provide the low angular momentum and
inflow required. 

\end{abstract}

\begin{keywords}
galaxies: kinematics and dynamics --- galaxies: individual: NGC\,891, NGC\,2403 --- galaxies: haloes --- galaxies: ISM --- galaxies: evolution --- ISM: kinematics and dynamics
\end{keywords}

\section{Introduction}

The evolution of spiral galaxies is influenced both internally by star
formation and externally by the environment.
Whether the former or the latter mechanism is dominant is difficult to 
determine.
Star formation is responsible for blowing gas outside the galactic
discs, perhaps polluting the Intergalactic Medium (IGM) with metals 
\citep[e.g.][]{car87}.
Outflows of gas from galactic discs with velocities of the order of
$100 \kms$ are observed both in the neutral
\citep[e.g.][]{kam91} and the ionised phase \citep[e.g.][]{fra04a}.
On the other hand, accretion of unpolluted material from the
IGM is predicted by chemical evolution models
to solve the so-called G-Dwarf problem in the Milky Way
\citep[e.g.][]{roc96}. 
The low metallicities of some of the High Velocity Clouds (HVCs) \citep{wak99}
suggest that cold material is indeed accreted by spiral galaxies like
our own \citep{Oort70}.

In the last years, the study of the connection between galactic
discs and haloes has flourished due to the collection of new
high-sensitivity data at different wavelengths.
Deep \ha\ observations have revealed the presence of extended layers  
of diffuse ionised gas (DIG) around edge-on spiral galaxies
\citep[e.g.][]{hoo99}.
X-ray observations with the new satellites
have shown the presence of hot gas at distances of tens of kpc from 
the plane of galactic discs \citep[e.g.][]{wan01}.
\hi\ observations of several spiral galaxies have revealed extended
thick layers or haloes surrounding the galactic discs \citep{swa97,
mat03}.
The kinematics of this extra-planar gas is characterized by:
1) a decrease in rotation velocity $v_{\phi}$ in the vertical
   direction \citep{swa97, fra05};
2) vertical motions from and towards the disc \citep{boo05};
3) a possible general radial inflow \citep{fra01}.

These studies show that the extra-planar gas is likely to be the
result of a complex exchange of gas between the disc to the halo regions.
In the galactic fountain scheme, gas is pushed up
by stellar activity, travels through the halo and eventually falls
back to the disc \citep{sha76,bre80}.
Recently, \citet{col02} presented a ballistic model of a galactic 
fountain and compared it with \ha\ long-slit observations of two
edge-on galaxies: NGC\,5775 and NGC\,891.
They found a disagreement, especially for NGC\,891, between the
rotational velocities above the plane predicted by their model and
those derived from the data.
Other studies have concentrated on the modelling of the
multi-phase ISM \citep{ros95, avi00}, the cooling of the hot coronal 
gas \citep{cor88} and the effect of turbulence \citep{str99}. 

Some authors have tried to model the extra-planar gas as a
stationary medium in hydrostatic equilibrium \citep{ben02, barn05}.
This approach
leads to solutions that can reproduce
the observed gradient in the rotation velocity above the plane if
the pressure of the medium does not depend on the density alone
\citep{barn05}.
However, the temperatures of the extra-planar (hydrostatic) gas are
of the order of 10$^5$ K and it is unclear how this medium can
be related to cold neutral gas. 

Here, we present a new model for the neutral extra-planar gas as the
result of a continuous galactic fountain flow.
We use a different strategy from the previous attempts mentioned
above.
We directly compare our dynamical models with \hi\ observations
(data cubes) by 
producing, as an output of the model, a {\it
pseudo}-data cube with the same total flux and resolution as the
data. 
The direct comparison with the observations is crucial because it
removes the intermediate steps of the data analysis such as the
derivation of rotation curves of the extra-planar gas or the
separation of the halo gas from the disc gas \citep[see][]{fra02} that
can introduce significant errors.

We apply this model to the observations of two spiral galaxies that 
currently provide the best examples of detection of extra-planar gas:
the edge-on galaxy NGC\,891 (Section \ref{s_n891}) and
NGC\,2403, which is seen at $63\de$ along the line of sight (Section
\ref{s_n2403}).
The study of galaxies viewed at different inclination angles is
essential to remove projection effects and obtain the full three-dimensional picture
of the distribution and kinematics of the extra-planar gas.
The data for an edge-on galaxy like NGC\,891 give information on the
extent of the gaseous halo and the rotation velocity of the gas
\citep{swa97}.  
In addition, 
the study of a galaxy at intermediate inclination like NGC\,2403
provides information on non-circular (inflow/outflow) motions 
of the extra-planar gas \citep{fra01}.

\section{The model} 
\label{s_model}

We have modelled the extra-planar gas in a spiral galaxy 
as a continuous flow of material from the disc to the halo.
Individual gas clouds are modelled as non-interacting
particles that are pushed up from the plane of the disc. 
The trajectory of each particle is integrated in the galactic
potential until it falls back to the disc.
Runs of the models were also performed in which the particles
cross the plane once but are stopped on their second passage.
Positions and velocities along these trajectories are projected along
the line of sight at each time interval, producing a {\it pseudo-cube} 
of sky positions and velocities to be compared with the \hi\ data cubes.

\subsection{The potential} 
\label{s_potential}

The galactic potential and forces were evaluated numerically on a grid in
the $(R,z)$ meridional plane, with cells $0.1\kpc$ on an edge close to the
plane and $0.5\kpc$ on a side further out.  The volume densities of both
the stellar and the gaseous components are given by
\begin{equation}
\rho_{\rm disc}(R,z) = \rho_0 e^{ - {R}/{R_{\rm d}} } \zeta
\left( \frac{z}{h_{\rm z}} \right),
\label{eq_rhodisc}
\end{equation}
 where $\rho_0$ is the central density,
$R_{\rm d}$ and $h_{\rm z}$ are the scale length and scale height of
the disc, and  
$\zeta(z)$ is the vertical density distribution, either an exponential
or a sech$^2$ law.
We assume that the scale height is independent of $R$.

The first part of the r.h.s.\ of eq.\ 1 is proportional to the
surface density of a razor-thin exponential disc.
 By modelling the disc as an infinitely
flattened homoeoid, one can show \citep{cud93} that its potential is
\begin{equation}
\Phi_{\rm thin}(R,z) = - \frac{4 G \Sigma_0}{R_{\rm d}} \int_{0}^{\infty}
\!\!\!{\rm d}a\, {\rm arcsin} \left( \frac{2a}{S_{+} + S_{-}} \right) a
K_{\rm 0}\left(\frac {a}{R_{\rm d}}\right),
\label{eq_phithin}
\end{equation}
 where 
$\Sigma_0$ is the central surface brightness,
$K_{\rm 0}$ is the modified Bessel function of zeroeth order, 
and 
$S_{\pm} \equiv \sqrt{z^2 + (a \pm R)^2}.$
The potential of the thick exponential disc (eq.\ \ref{eq_rhodisc})
is
\begin{equation}
\Phi_{\rm thick}(R,z) = \int_{- \infty}^{\infty} {\rm d}z' \zeta(z')
\Phi_{\rm thin}(R,z),
\label{eq_phithick}
\end{equation}
where
$\Phi_{\rm thin}(R,z)$ is as in eq.\ \ref{eq_phithin} but with 
 \[
S_{\pm} \equiv \sqrt{(z-z')^2 + (a \pm R)^2}.
\]

The ellipsoidal components (bulge and dark matter halo)
have been modelled using a double power law density profile
\citep{deh98}:
\begin{equation}
\rho_{\rm dpl}(R,z)=\rho_{\rm 0, dpl} \left(\frac{m}{a} \right)^{-\gamma} 
\left(1+\frac{m}{a} \right)^{\gamma-\beta},
\label{eq_rhodpl}
\end{equation}
 where $\rho_{\rm 0, dpl}$ is the central density,
$a$ is the scale length,
$\gamma$ and $\beta$ are the inner and outer slopes respectively,
and $m \equiv \sqrt{R^2+{z^2}/{q^2}}$
with $q$ the axis ratio, related to the eccentricity $e$ by 
$q=\sqrt{1-e^2}$.

The potential and forces of the ellipsoidal components are calculated
using formulae (2-84b) and (2-88) in \citet{bin87} with the
substitution of the density $\rho$ with that shown in our eq.\
\ref{eq_rhodpl}. 

\begin{figure}
\begin{center}
  \includegraphics[width=200pt]{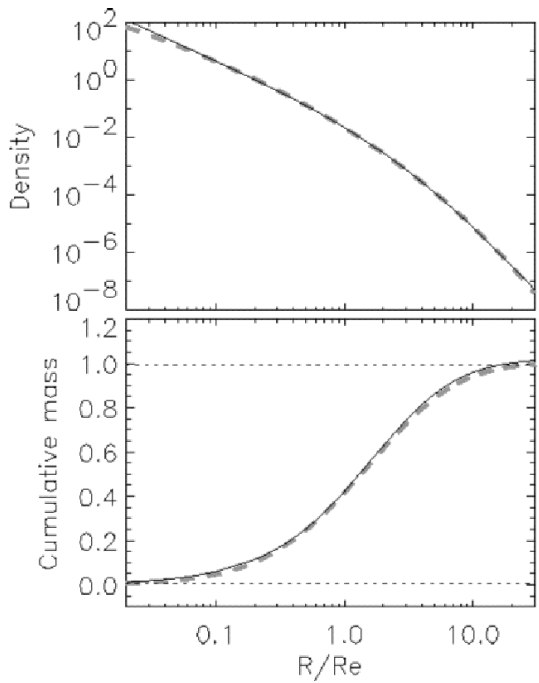}
  \caption{Fit to the volume density corresponding to a projected
  R$^{1/4}$ profile (dashed lines) with the double power law profile
  of eq.\ \ref{eq_rhodpl} (solid line).
  The dotted straight lines in the right panel mark the levels 0.5\%
  and 99.5\% of the total mass. 
\label{f_fitdev}
}
\end{center}
\end{figure}

The double-power-law profile allows us to model dark matter (DM) haloes with
different shapes such as pseudo-isothermal, $\gamma=0$ and $\beta=2$; NFW
profile, $\gamma=1$ and $\beta=3$ \citep{nav97}; Hernquist profile,
$\gamma=1$ and $\beta=4$ \citep{her90}.  We have also used these profiles to
model bulges that satisfy de Vaucouleurs' $R^{1/4}$-law \citep{dev48}. The
associated luminosity density is given by \citet{you76}, and we have fitted
this density with the double power law profile selecting a radial range that
encloses 99\% of the light.  The best fit, for $\gamma$ and $\beta$ integer
numbers, gives $\gamma=2$, $\beta=5$, $a=3.28R_\e$, where $R_\e$ is the
effective radius of the $R^{1/4}$-law.  Solid lines in Fig.~\ref{f_fitdev}
show this fit, while the dashed grey lines show the $R^{1/4}$ profile.

For each galaxy we have constructed three mass models: 1) maximum-light
(stellar disc + bulge); {2) maximum spherical DM halo; 3) maximum flat DM
halo.}  In the first mass model, the stellar disc and bulge contribute
maximally to the rotation curve; in the second and third models, the stellar
mass-to-light ratio is set to the {\it minimum\/} plausible value (see
Sections \ref{s_massmodel891} and \ref{s_massmodel2403}).  The parameters of
the dark matter haloes are then tuned to  reproduce the rotation
curve.  The eccentricity is $e=0$ in model 2 and $e=0.95$ (axis ratio
$q\sim0.3$) in model 3.

\subsection{Initial conditions} 
\label{s_conditions}

The orbits of the particles are integrated in the $(R,z)$ plane 
and then uniformly distributed in the azimuthal angle $\phi$. Thus each
orbit is used to follow an ensemble of clouds that are launched from
locations lying around a circle in the plane.
{The integration is performed in cartesian coordinates using the fourth-order
Runge-Kutta method}.
At each interval $\delta t$, the positions and the velocities of the particles
are projected along the line of sight.
{We used a variable time step with upper value
$\delta t=3\times 10^5$ yr.}

\begin{figure}
\begin{center}
  \includegraphics[width=240pt]{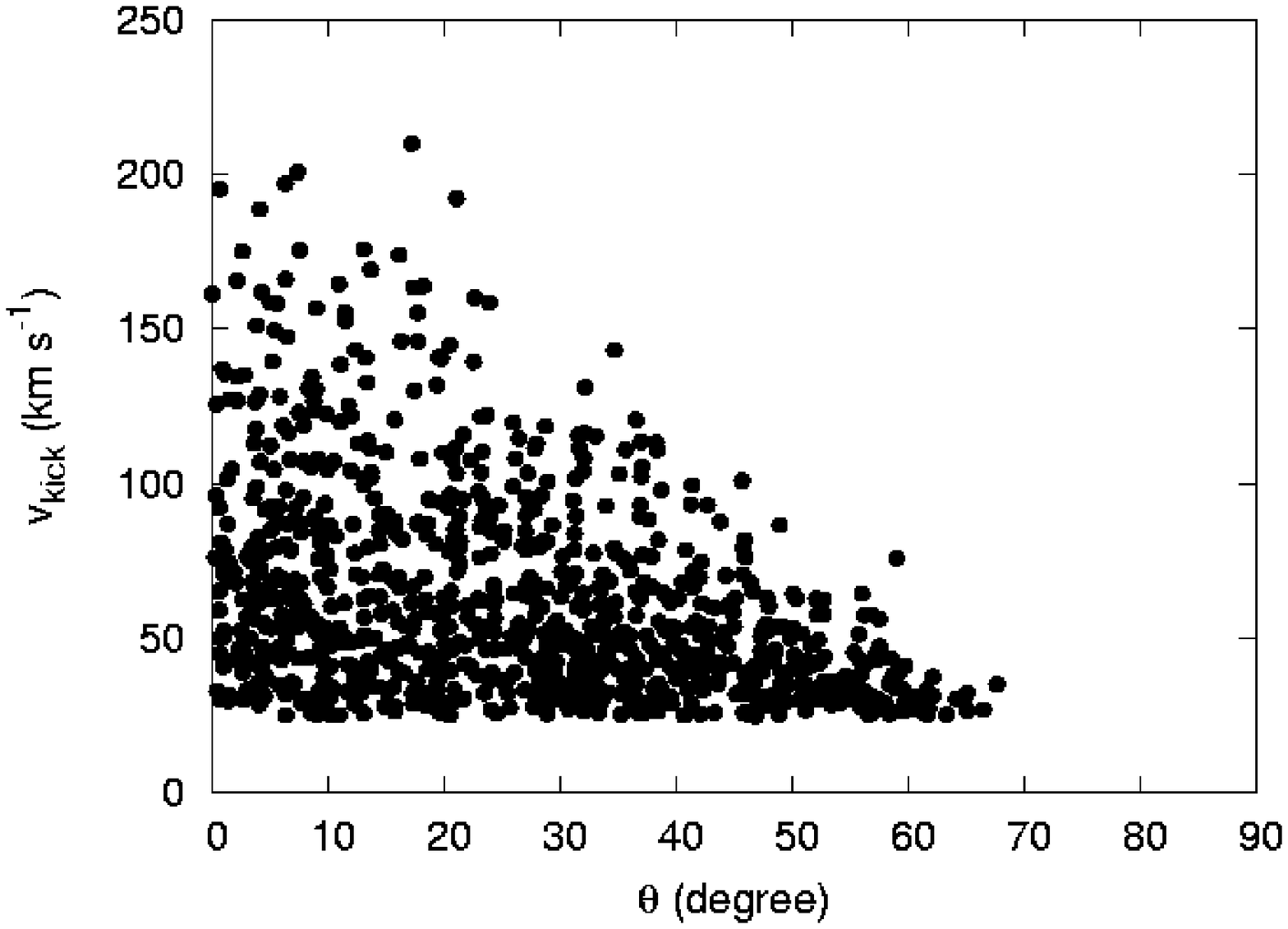}
  \includegraphics[width=240pt]{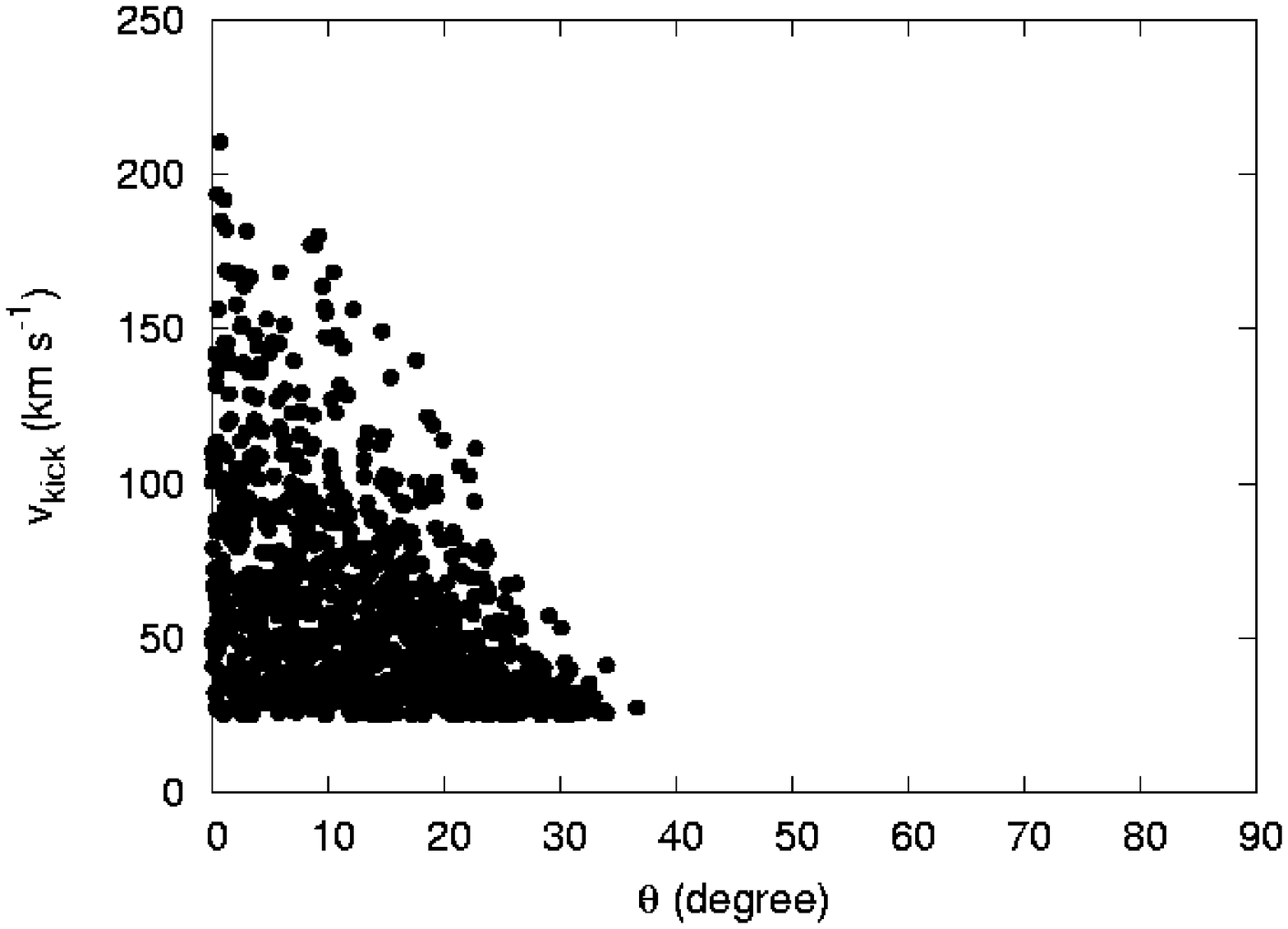}
  \caption{Distribution of kick velocities for two values of
  $\Gamma=2$ (top) and $\Gamma=10$ (bottom) in eq.\ \ref{eq_p}. The 
  two distributions are Gaussian with $h_{\rm v} = 75 \kms$. 
\label{f_gamma}
}
\end{center}
\end{figure}

Initially, the particles are considered to be gyrating around circular
orbits in the plane $z=0$ in such a way that they have isotropic 
random velocities with dispersion $\sigma_{\rm ran}=7\kms$.  Then the
particle receives a kick velocity in a direction that is chosen at random
from an axisymmetric distribution. The magnitude of $\vec{v}$ has
probability distribution
 \begin{equation}
P(\vec{v}) \propto \xi(v, \theta),
\label{eq_p}
\end{equation}
 where 
$\theta$ is the angle between the direction of the kick and the
normal to the galactic plane. An natural choice for $\xi$ is
 \begin{equation}
\xi_{\rm G}(v,\theta) = \exp \left( -\frac{v^2}{2 h_{\rm
v}^2 \cos ^{2 \Gamma} \theta} \right),
\label{eq_xiG}
\end{equation}
 where  $\Gamma$ and the characteristic kick velocity
$h_{\rm v}$ are parameters to be determined. The highest velocities
are confined to a cone around the normal to the galactic plane, and the
width of this cone is set by the parameter $\Gamma$.  The functional shape
of $\xi$ determines the
thickness of the gas halo and has to be constrained by the observations
(Section \ref{s_constraints}).  In order to avoid numerous low-velocity
integrations, we have included a lower cut-off for kick velocities $v_{\rm
thres}\simeq25\kms$. 
Fig.~\ref{f_gamma} shows the resulting distributions of kick
velocities as a function of $\theta$ with $h_{\rm v}=75 \kms$ and two choices of
$\Gamma$ ($2$, top panel and $10$, bottom panel). 
In section \ref{s_constraints} we show how to constrain the value of
$\Gamma$ from the data.
It turns out that high values ($\Gamma \gsim5$) are required
leading to a fairly collimated flux.

Stochastic acceleration processes sometimes produce a  tail of particles
that extends to large $v$, which is absent if $\xi$ is taken to be the
Gaussian function (\ref{eq_xiG}). Therefore we examine the possibility that
$\xi$ tends to a power law at large  $v$. Specifically, we present results
for the choice
 \begin{equation}
\xi_{\rm PL}(v, \theta)= 
\left\{
\begin{array}{ll}
{\displaystyle 
\left(1+\frac{v}{v_* \cos ^{\Gamma} \theta} \right)^{- \alpha_{\rm v}}
} &v \leq v_{\rm cut}\\
0 & v> v_{\rm cut},\\
\end{array}\right.
\label{eq_xiPL}
\end{equation}
 where $v_*$ is a characteristic velocity that we identify with the velocity
dispersion of the ISM, $v_*=7 \kms$, and the slope $\alpha_{\rm v}$ and the
cut-off velocity $v_{\rm cut}$ are free parameters.

We have also experimented with forms of $P(\vec{v})$ that are not functions
of $v/\cos^\Gamma\theta$, as is the case in eqs  (\ref{eq_xiG})
and (\ref{eq_xiPL}). We find, for example, that when $P$ is of the form
$P(\vec{v})=\xi(v) \cos^{\Gamma}\theta$,
the data require even larger values of $\Gamma$ than those just discussed.

\subsection{Outflow rate and halo mass} 

In our model, the flow of material escaping the disc is considered 
constant throughout the life of the galaxy. 
This is a reasonable assumption for a normal spiral like the Milky Way 
given the regularity of the star formation history
\citep[e.g.][]{roc00}.  
Therefore, at all times, there will be a constant number of particles
(mass of gas) in the halo.
Of these, the number launched between $R$ and $R+ \Delta R$ and with kick
velocities between $\vec{v}$ and $\vec{v}+\Delta \vec{v}$ will be: 
 \begin{equation}
\Delta n_{\rm halo}(R,\vec{v}) =2 \pi R \dot n_{\rm
out}(R,\vec{v}) \tau(R,\vec{v}) \Delta R \Delta \vec{v},
\label{eq_n_halo}
\end{equation}
 where $\dot n_{\rm out}(R,\vec{v})$ is number of particles (outflow rate)
generated per unit time and area at radius $R$ with kick velocity $\vec{v}$,
and $\tau(R,\vec{v})$ is the travel time for these particles.  We can
re-write the outflow rate
 \begin{equation}
\dot n_{\rm out}(R,\vec{v}) = \dot n_{\rm out}(R) \times P(\vec{v}),
\label{eq_n_out}
\end{equation}
 where 
$P(\vec{v})$ is the distribution in eq.\ \ref{eq_p} normalized such that 
$\int_{v_{\rm thres}}^{\infty} P(\vec{v}) d\vec{v} = 1$.

Since the causes of gas outflows are, presumably, strong stellar winds 
and supernova explosions, we  assume that the outflow rate is 
proportional to the star formation rate (SFR). 
The latter can be taken proportional to the gas surface density
$\Sigma_{\rm g}(R)$ to a power $\alpha_{\rm SF}$ (Schmidt law)
and therefore:
\begin{equation}
\dot n_{\rm out}(R) \approx H~ [\Sigma_{\rm g}(R)]^{\alpha_{\rm SF}},
\label{eq_schmidt}
\end{equation}
 where $\alpha_{\rm SF}\sim$1.3 \citep{ken89} and
$H$ is a constant that ultimately represents 
the efficiency of the transfer of kinetic energy from the stellar
activity to the halo gas.
We can relate this parameter to the total mass of the halo gas. 
If we integrate eq. \ref{eq_n_halo} over the whole galaxy, we find
that the total mass of the gaseous halo is
 \begin{equation}
\begin{array}{ll}
M_{\rm halo}= n_{\rm halo} \times m_{\rm p}\\ 
~\\
~~~~~~~= 2 \pi m_{\rm p} H
\int_{0}^{R_{\rm cut}}d R\, R \Sigma_{\rm g}^{\alpha_{\rm SF}}
\int_{v_{\rm thres}}^{\infty} d\vec{v} \tau(R,\vec{v}) f(\vec{v}),
\end{array}
\label{eq_m_halo}
\end{equation}
 where $m_{\rm p}$ is the mass of the single particle and
R$_{\rm cut}$ is the cut-off radius of the SF activity in the disc.  
Therefore, for any given $M_{\rm halo}$ we obtain a value for $m_{\rm
p}H$ and an outflow rate (eq.\ \ref{eq_schmidt}) in proper units.

\subsection{Comparison to the observations}

The strategy of this study is to compare our models with \hi\
observations of spiral galaxies showing the presence of extra-planar
gas. 
In the model, at each time interval $dt$, positions and velocities of
the particles are projected along the line of sight for a specified
inclination of the galaxy.
From the projected velocities we construct a model cube of particles 
that can be compared with the \hi\ data cube.
The total flux in the model cube is normalized to the total \hi\ mass 
of the galaxy.
The spatial and velocity resolutions are smoothed to those of the
data.
A disc
component is added to the output of the model with a density profile
derived from the data and a mass $M_{\rm disc} = M_{\rm tot} - M_{\rm 
halo}$, where $M_{\rm tot}$ is the total \hi\ mass of the galaxy.
{The number of particles used in the simulation is on average 
$\sim 100$ per beam.
This, in the case of NGC\,891, implies a typical particle 
mass of $\sim 10^4 \mo$, two orders of magnitude below the observational 
detection limit.
Therefore, our model is smooth and different simulations do not 
show significant fluctuations with respect to each other.
}

The surface density $\Sigma_{\rm g}(R)$ in eq.\ \ref{eq_schmidt} has been
derived, for NGC\,891, considering both the atomic and molecular gas.  The
distribution of neutral gas was derived from the \hi\ data whilst that of
H$_2$ was taken from CO observations \citep{sof93}.  The fit of the total
gas density with an exponential profile gave $R_{\rm gas}$=5.0 $\pm$ 0.5 kpc
and the total mass M$_{\rm HI+H_2}=7.9\times 10^9\mo$, in good agreement
with the total mass that one gets using the \hi\ and CO fluxes: $M_{\rm
tot}=M_{\rm HI}+M_{\rm H_2}=4.0+(3.2- 4.1)\times\mo$, the latter values
referring to two different choices for the CO/H$_2$ conversion factor
\citep{sof93}.  For NGC\,2403, we have consider only atomic gas, being the
dominant component and used the exponential fit to the \hi\ surface density
given in \citet{fra02}.

\section{Application to NGC\,891} 
\label{s_n891}

The (Sb) spiral galaxy NGC\, 891 is one of the best known and
extensively studied nearby edge-on galaxies.
We assumed a distance of $9.5\,$Mpc \citep{vdk81b}, which leads to a
luminosity $L_{\rm B}\simeq2.6\times10^{10} \lo$.
The disc of NGC\,891 shows intensive star formation at a rate SFR
$\simeq 3.8 \moyr$ \citep{pop04}. 
The halo region has been studied at various wavelengths and shows a
variety of components from radio continuum \citep{all78} to hot X-ray
emitting diffuse gas \citep{bre94}.
NGC\,891 is also considered to be very similar to the Milky Way
\citep{vdk84}.

NGC\,891 has been studied in HI several times with ever increasing
sensitivity \citep[e.g.][]{san79, rup91, swa97} and most recently with very
deep observations (with more than 200 hours of integration) 
obtained with the Westerbork
Synthesis Radio Telescope (WSRT) \citep{oos05}.
We use here these new data at 28$''$ (1.3 kpc) resolution.  The total \hi\ map,
rotated $67\de$ counter-clockwise, is shown in Fig.~\ref{f_models891_tot_1}
(upper left panel).  The distribution of neutral gas in the disc of NGC\,891
is not symmetric, being more extended in the South-West side (on the right
of the rotated image).  The   location of this
extension is not known (it is probably not
in the line of nodes), so in deriving the rotation curve and studying
the kinematics, it is often neglected.  These new WSRT observations show
very clearly the presence of extended extra-planar emission, much more
extended that in the previous observations \citep{swa97}.  As shown both by
\citet{swa97} and \citet{fra05}, this emission cannot be explained by
projection  of either a flare or a line-of-sight warp, but it has to come
from a thick component of neutral gas surrounding the thin disc and lagging
behind it in rotation.  The extent and the lag of the halo gas are the key
constraints on our model.

\subsection{Mass models for NGC\,891} 
\label{s_massmodel891}

\begin{figure}
\begin{center}
  \includegraphics[width=240pt]{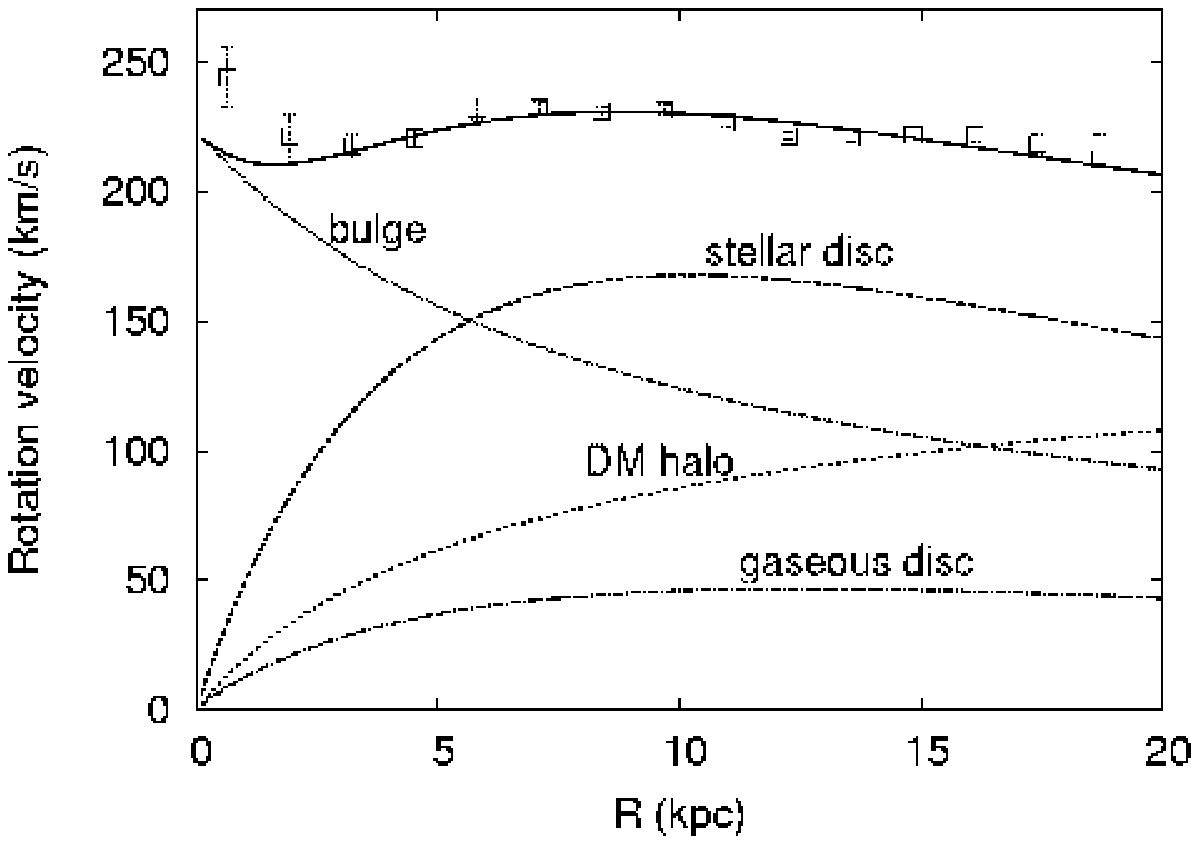}
  \includegraphics[width=240pt]{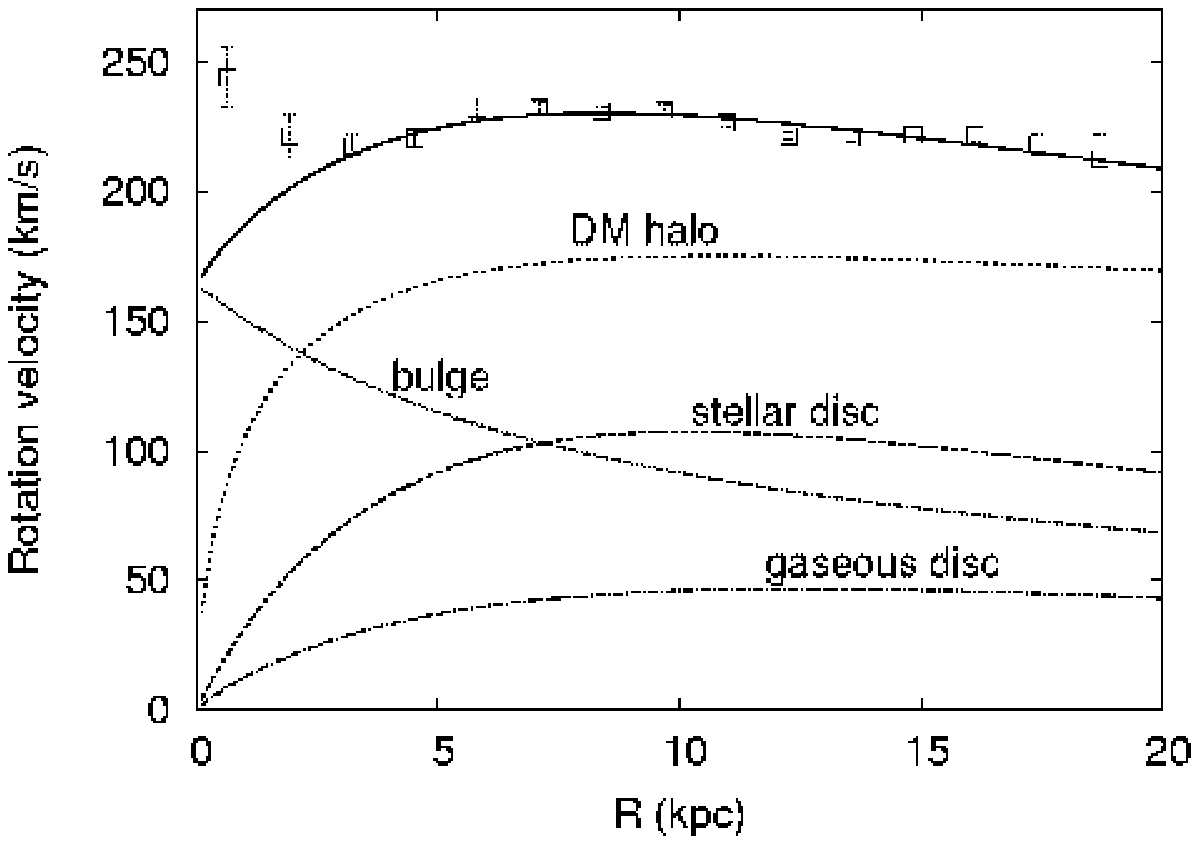}
  \caption{Two fits to the \hi\ rotation curve of NGC\,891 (squares)
  obtained with the maximum light (top panel) and maximum halo (bottom
  panel) models. 
  The two inner points of the rotation curve have not been included in
  the fit due to uncertainties on their interpretation (see text). 
  \label{f_massmodels891}
}
\end{center}
\end{figure}

The rotation curve of NGC\,891 (Fig.~\ref{f_massmodels891}) has been
derived using an envelope tracing technique \citep[e.g][]{san79} applied to
the north-west side of the galaxy (details in Fraternali, in preparation).
The velocity rises very steeply in the inner parts of the galaxy ($R<
2\kpc$), this phenomenon may be evidence of a fast rotating inner ring or,
perhaps non-circular motions related to an inner bar \cite[see][]{gar95}.
Due to these uncertainties in interpretation of the inner points of the
rotation curve we have excluded them in our fitting. 

\begin{table*}
 \centering
   \caption{Mass models for NGC\,891} 
   \label{t_massmodels891}
   \begin{tabular}{@{}lccccccccccc@{}}
     \hline
     Model & $(M/L)_{\rm disc}$ & $R_{\rm d}$ & $h_{\rm z}$ &
     $(M/L)_{\rm bulge}$ & $R_e$ & $e_{\rm bulge}$ & $\rho_{\rm
       0,DM}$ & $a$ & $\gamma$ & $\beta$ & $e_{\rm DM}$ \\ 
     & $(\!\mo/\loB)$ & (kpc) & (kpc) & $(\!\mo/\loB)$ & (kpc) & & $(\!\mo 
     \kpc^{-3})$& (kpc) & & &\\ 
     \hline
     maximum light & 3.7 & 4.4 & 1.05 & 6.3 & 3.3 & 0.8 &
     $3.7\times10^7$ & 4.0 & 0 & 2 & 0 \\
     maximum round DM halo & 1.5 &  4.4 & 1.05 & 3.5 & 3.3 & 0.8 &
     $1.05\times10^8$ & 5.0 & 1 & 3 & 0 \\
     maximum flat DM halo & 1.5  & 4.4 & 1.05 & 3.5 & 3.3 & 0.8 & 
     $1.76\times10^8$ & 6.0 & 1 & 3 & 0.95 \\
     \hline
   \end{tabular}
\end{table*}

We decomposed the potential of NGC\,891 into four components: two
discs (stellar and gaseous), the potentials of a bulge and a DM halo. 
For the stellar component we have used the photometry and disc-bulge
separation of \cite{sha89}, who
used the same optical data as \cite{vdk81b}. 
They found no significant difference in accuracy between fits with two
or three components (the latter with thin $+$ thick discs).
Therefore we considered just two components, a relatively thick disc
($h_{\rm z}\simeq1\kpc$) and a  flattened $R^{1/4}$-law bulge.
This, as \citet{sha89} clearly state, is not a perfect description of
the light distribution in NGC\,891 but it is accurate enough for our
purposes. 
Indeed (see Section \ref{s_effects891}) our main conclusions do not
depend on the shape of the potential.

Fig.~\ref{f_massmodels891} shows the fits to the rotation curve of
NGC\,891 obtained with two of the adopted mass models.  The first (top
panel) is a maximum light (disc+bulge) model, in which the highest possible
mass-to-light ratio is assigned to the luminosity density.  The $M/L_{\rm
B}$ ratios obtained for the stellar disc and the bulge are, 3.7
and 6.3,  respectively. The disc and bulge
are dominant all the way out to the last point of the rotation curve.  The
dark-matter halo is pseudo-isothermal, having inner and outer slopes of
$\gamma=0$ and $\beta=2$, respectively (see eq.\ \ref{eq_rhodpl}).

The bottom panel of Fig.~\ref{f_massmodels891} shows the fit to the
rotation curve with a  maximum-halo model. 
In this case the $M/L_{\rm B}$ ratios were fixed to 1.5 and 3.5 respectively 
for disc and bulge. 
We chose these values as the minimum allowed for two stellar 
populations of mean ages of about 2 and 4-5 Gyr, for disc and bulge 
respectively \citep{cha96}.
In order to reproduce the rotation curve in this case we need to use a 
more peaked DM halo of a NFW type ($\gamma=$1, $\beta=$3).
The DM halo dominates everywhere outside $R=2\kpc$. 
The model shown in Fig.~\ref{f_massmodels891} is for a round halo ($e=0$). 
We also constructed a model for a flat  halo ($e=0.95$, $q\simeq0.3$),
where the fit to the rotation curve is essentially identical to that
for a round halo.
The parameters of the mass models of NGC\,891 are reported in Table
\ref{t_massmodels891}.

\subsection{Model constraints} 
\label{s_constraints}

\begin{figure*}
  \includegraphics[width=400pt]{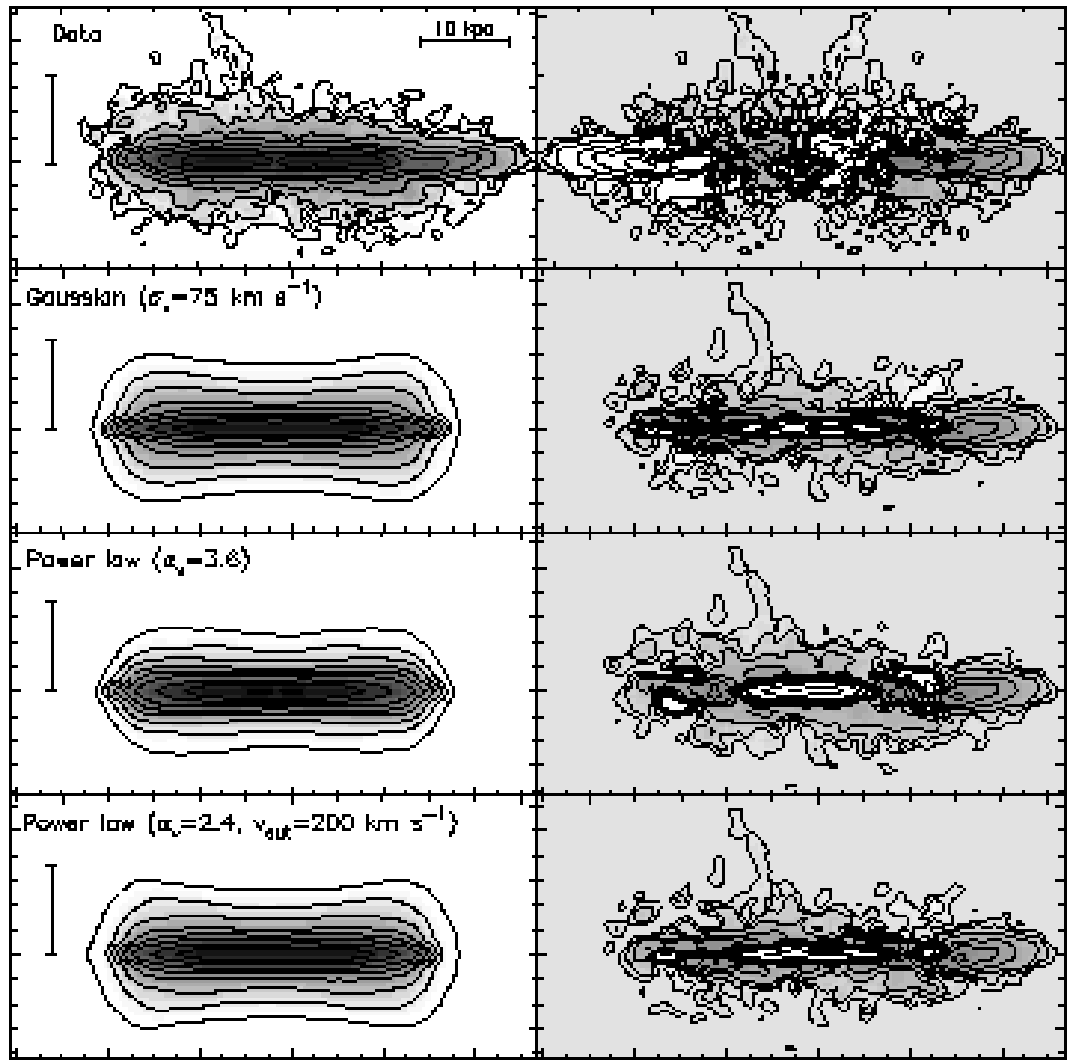}
  \caption{Comparison between the (rotated) total \hi\ map of NGC\,891
    (upper left panel) \citep{oos05} and the results from three dynamical 
models. 
    The models all have a maximum light potential but differ on the shape
    of the distribution function of the kick velocities (eq.\ \ref{eq_p}).
    The right panels show the residuals of the subtraction of the model
    total maps from the total \hi\ map.
    The upper right panel shows the results of subtracting a minor-axis
    mirrored \hi\ map from the original data. 
    In the left panels the contour levels are: 0.2, 0.45, 1, 2.5, 5, 10,
    25, $50 \mopc$. 
    {In the right panels the grey background indicates the zero level, 
darker colours imply that the measured intensity is greater than that
predicted by the model. Contour levels are the same as in the left panels.
}
    \label{f_models891_tot_1}
  }
\end{figure*}

The most striking evidence for extra-planar gas in 
NGC\,891 is given by the total HI map (Fig.~\ref{f_models891_tot_1},
upper left panel; see also \citet{fra05}) and
our first step is to reproduce the observed vertical
distribution.
Hence, we have explored the parameter space to obtain the right amount of
high-latitude gas. 
The parameters of our model that have a critical impact are:
1) the functional form of $P(\vec{v})$ (eq.\ \ref{eq_p}) with its
   parameters ($h_{\rm v}$, $\alpha_v$ and $v_{\rm cut}$); 2) the mass of
   the halo gas (eq.\ \ref{eq_m_halo}); 3) the cut-off radius of
   star formation (eq.\ \ref{eq_m_halo}). 

To constrain these parameters we have used an iterative method. We have
constructed total maps with different values of these parameters, subtracted
them from the total \hi\ map of NGC\,891 and minimized the residuals in the
{halo region only (at  $|z|\sim1\kpc$)}.
For each of the three functional forms of $P(\vec{v})$,
the lower three panels in the left column of Fig.~\ref{f_models891_tot_1}
show the best fits to total maps that can be obtained by adjusting the free
parameters.  The parameters of these models are given in Table
\ref{t_models891}. In each case the cut-off radius for SF is $R_{\rm
cut}=16\kpc$ and the potential is that of the maximum-light model. Hence the
models differ only in the distribution of kick velocities.

The upper right panel of Fig.~\ref{f_models891_tot_1} shows the residuals
obtained just mirroring the data about the minor axis and subtracting it
from the original total \hi\ map.  Clearly, no axisymmetric model could
produce smaller residuals than these.  The lower three panels of the right
column show the residuals produced by the model whose total \hi\ map is shown
on the left.  In
the plane there are systematic differences between data and models; these
residuals reflect both
inhomogeneities in the plane and uncertainties in
the shape of the  radial \hi\ profile.   Inevitably, the model does not
match the outer part of the disc on the right (originally South-West) side
of the galaxy.

Our models tend to produce a flaring shape of the total map because
particles launched from larger radii can reach higher distances from the
plane.  Below the plane (South-East in the original orientation), the data
show a hint of a flare, but the effect is weaker in the data than in most
models.  This flare can be reduced by increasing the parameter $\alpha_{\rm
SF}$ in eq.\ \ref{eq_schmidt}. The optimum value was found to be close to
1.3. However, overall the residuals are insensitive to $\alpha_{\rm SF}$, so
for NGC\,891 we henceforth simply set $\alpha_{\rm SF}=1.3$.  In Section
\ref{s_forbidden} we vary $\alpha_{\rm SF}$ in an attempt to
account for gas at very peculiar velocities near the centre of NGC\,2403. 

\begin{figure}
  \includegraphics[width=240pt]{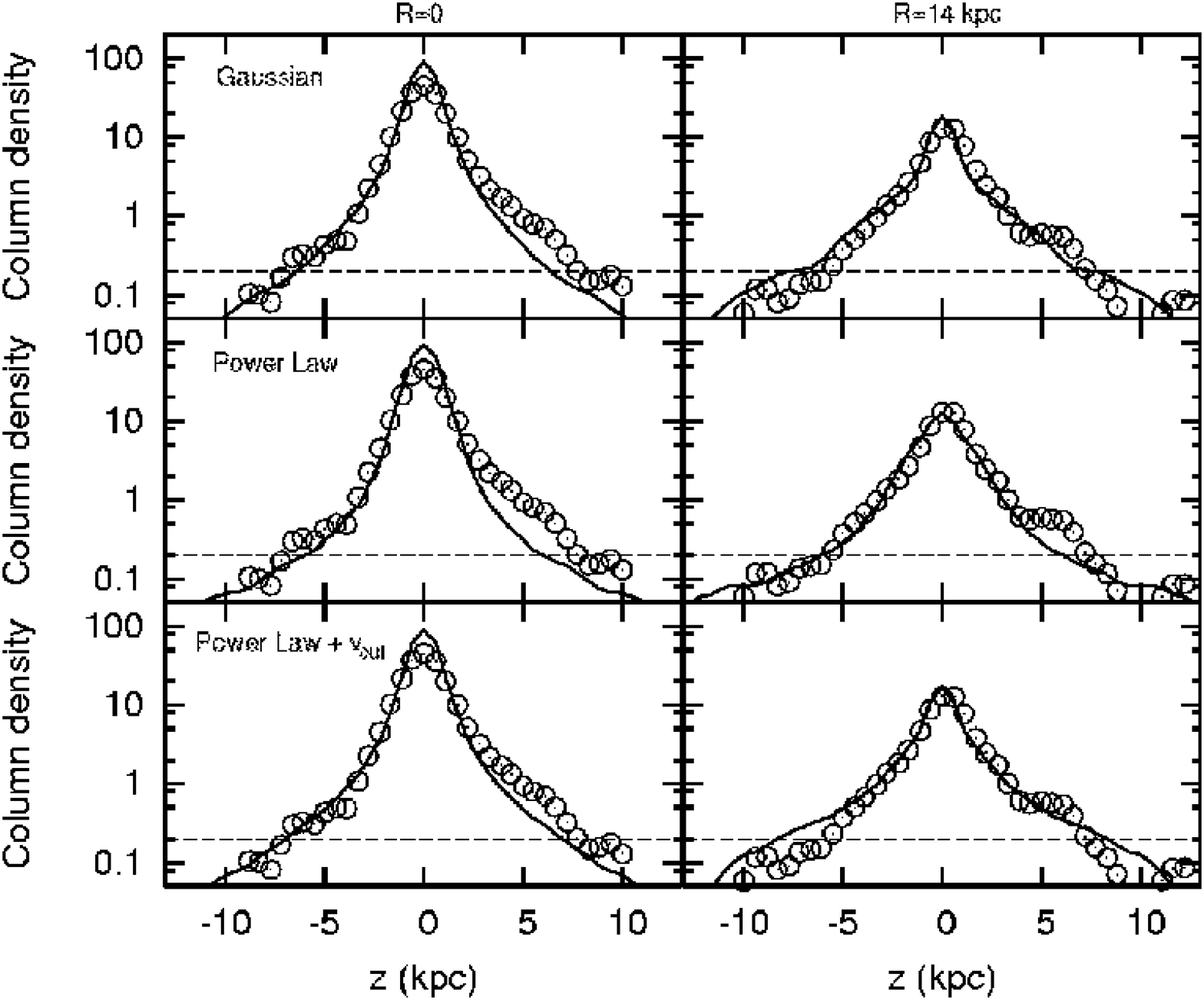}
  \caption{Comparison between the vertical density profiles for
  NGC\,891 (circles) and the three models (solid lines) shown in
  Fig.~\ref{f_models891_tot_1} (from top: Gaussian, PL, PL$+$v$_{\rm cut}$). 
  The left panels are taken at the centre of the galaxy and the right
  panels at about $14\kpc$ towards left (North-East in the original 
  orientation).
  The left parts of the plots correspond to the lower parts in the
  maps in Fig.~\ref{f_models891_tot_1}. 
  The column densities are in $\mopc$ and the dashed lines show the
  level of the last contour in the total 
  \hi\ map in Fig.~\ref{f_models891_tot_1}.
\label{f_zprofiles_1}
}
\end{figure}

In Fig.~\ref{f_zprofiles_1} we plot density profiles perpendicular to the
plane, the vertical ordering of the models being as
in Fig.~\ref{f_models891_tot_1}.  The left-hand profiles are along the minor
axis, while the right panels are along the parallel line that cuts the major
axis $14\kpc$ to the left of the centre in Fig.~\ref{f_models891_tot_1}.
Circles show the data, and the column-density scale is in $\!\mopc$.  The
dashed lines show the last contour level in the total \hi\ map. 

\begin{table}
 \centering
 \begin{minipage}{140mm}
   \caption{Parameters of the maximum light models for NGC\,891}
   \label{t_models891}
   \begin{tabular}{@{}lcccc@{}}
     \hline
     Model & $h_{\rm v}$ & $\alpha_{\rm v}$ & $M_{\rm halo}$ &
     $v_{\rm max}$ \\
     & $(\!\kms)$ &  & $(10^9 \mo)$ & $(\!\kms)$ \\
     \hline
     Gaussian & 75 & - & 2.0 & $\infty$ \\
     Power Law & - & 3.6 & 3.0 & $\infty$ \\
     Power Law + v$_{\rm cut}$ & - & 2.4 & 2.3 & 200 \\
     \hline
   \end{tabular}
 \end{minipage}
\end{table}

The best results are obtained with the Gaussian and the Power Law (PL)$+$v$_{\rm cut}$ models.  
The PL model without a velocity cut-off does not produce much
extra-planar gas above the detection limit, instead it produces a long tail
of emission below this limit (partially visible in
Fig.~\ref{f_zprofiles_1}).  This is why the total map of the PL model
appears thinner than the others and there are substantial residuals.  Note
however that a lower value of $\alpha_v$ would have produced too much
material at very high latitude.
This shows that, except to produce the high latitude spur visible in
the total \hi\ map, there is no need for a
population of clouds with very high kick velocities (above $200 \kms$) that
may arise from an un-truncated PL distribution of velocities.
Therefore, in
the following models, we prefer to use the Gaussian distribution, which has
only one free parameter.  The characteristic kick velocity required to
reproduce the distribution of the halo gas is around $75 \kms$ ($\sim1/3$ of
the circular velocity in NGC\,891).  This number is well constrained ($\pm10
\kms$): for lower $h_{\rm v}$ the extent of the halo is not reproduced.

It is interesting to compare this characteristic value for the kick
velocities with the expansion velocities for superbubbles that are predicted
by hydrodynamical models.  In the classic paper by \citet{mac88}, in the
case of an exponential atmosphere and reasonable values for the other
parameters, the expansion velocities at heights above 300 pc (when the
supershell ``leaves'' the plane) are expected to be of $v_{\rm exp} \gsim 50
\kms$ in good agreement with our findings.  Moreover, observations of
vertical motions in \hi\ gas give very similar values.  A well known example
is the expanding supershell found in M\,101 \citep{kam91} for which a
line-of-sight expansion velocity of $\sim50 \kms$ has been derived. 

\begin{figure}
  \includegraphics[width=240pt]{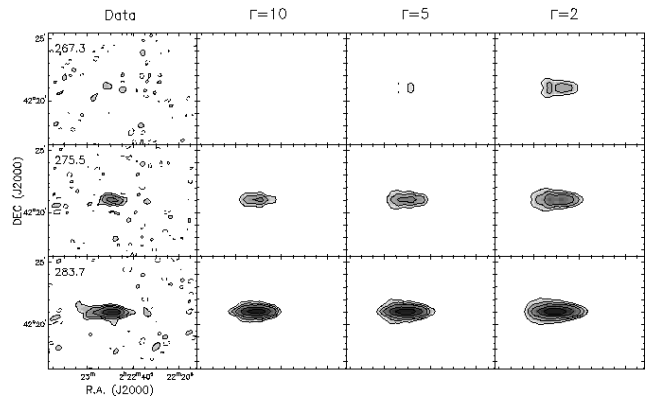}
  \caption{Comparison of three channel maps of NGC\,891 (first column)
  and those of three models with different opening angles for the kick
  velocities (parameter $\Gamma$ in eq.\ \ref{eq_p}).
{The first column shows the data with heliocentric radial velocities
  reported in the upper left corners ($V_{\rm sys}=528 \kms$).}  
The data are consistent with values of $\Gamma\gsim5$.
\label{f_models891_gamma}
}
\end{figure}

We now study the effect of the parameter $\Gamma$ in eq.\ \ref{eq_p}.  This
parameter affects the direction of the kick velocities.  The larger $\Gamma$
is, the closer the initial trajectories are to the normal to the plane
(Fig.~\ref{f_gamma}).  Fig.~\ref{f_models891_gamma} shows channel maps
(heliocentric velocities are shown in the upper left corner) for the data of
NGC\,891 and for three models with different values of $\Gamma$.  The three
models have the same parameters as the Gaussian model in Table
\ref{t_models891} and Fig.~\ref{f_models891_tot_1}.  The channel maps shown
here are those at extreme approaching velocities.  Varying the opening angle
produces different patterns especially for these channels.  As one can see
from this figure, if the cone of the kick velocities is too wide (low
$\Gamma$), particles end up populating channels that are not populated in
the data.  For $\Gamma\gsim5$ the channel maps are similar to those of the
data.  We have fixed $\Gamma=8$ for all our models.  {When $\Gamma \gsim5$,
$P(v)$ with $v\simeq h_{\rm v}$ falls by a factor two as $\theta$ increases
from $0$ to $15^\circ$. 
Thus, the data require a fairly highly collimated upward flow.}

\subsection{Dynamics of the extra-planar gas} 
\label{s_dynamics891}

\begin{figure*}
  \includegraphics[width=400pt]{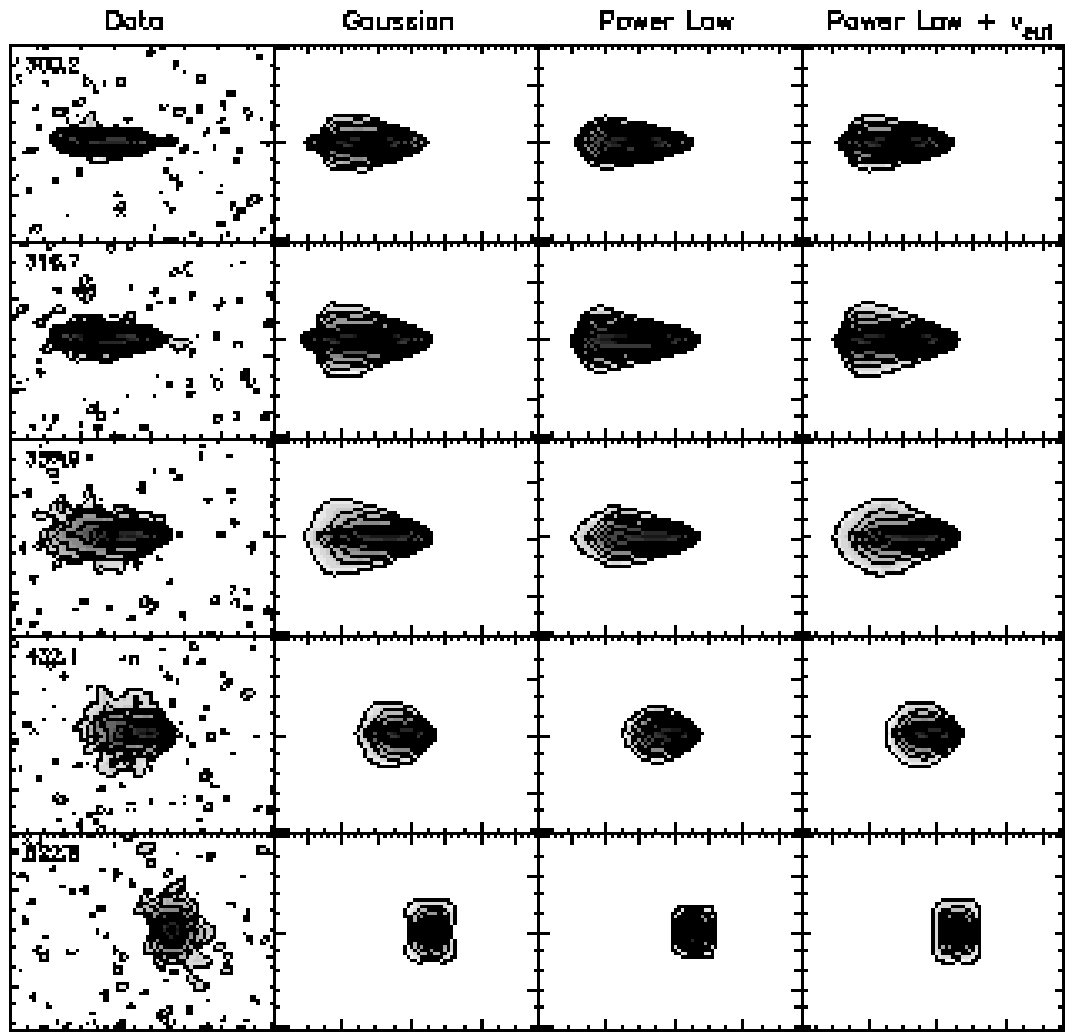}
  \caption{Comparison between 5 channel maps for NGC\,891 \citep{oos05}
    and those obtained with three dynamical models with different
    distribution of the kick velocities (Table \ref{t_models891}).
    The first column shows the data, heliocentric radial velocities
    are reported in the upper left corners. 
{The channel map in the bottom row is roughly at the systemic velocity 
($V_{\rm sys}=528 \kms$).
}
    Contour levels (for data and models) are: 0.45 (2$\sigma$), 1, 2,
    5, 10, 20, 50 mJy/beam. 
    \label{f_models891_chan_1}
  }
\end{figure*}

Extra-planar gas rotates more slowly than the gas in the plane.  This may be
seen in the channel maps.  Fig.~\ref{f_models891_chan_1} (leftmost column)
shows five representative channel maps in the approaching (North-East) side
of the galaxy, compared with the three models of Table \ref{t_models891}.
The emission of NGC\,891 becomes thinner the more one approaches high
rotation velocities (first rows), far from the systemic one ($v_{\rm
sys}=528 \kms$).  Although this effect can be produced also by other
phenomena, especially a line of sight warp, \citet{swa97} showed that the
shape of the channel maps is accurately reproduced only by a lagging halo.
The presence of a mild warp along the line of sight cannot be excluded
\citep{bec97} but it will play a minor role.  In our models we have assumed
that the galaxy is perfectly edge-on at each radius, so all
the emission outside the plane is produced by extra-planar gas. 

Although the models reproduce the basic shape of the channels, there are
systematic differences with the data.  In the data, extra-planar gas is more
concentrated to channels close to the systemic velocity (bottom row) and is
almost absent from the extreme channels (two first rows).  Even though the
model channel maps at $v_{\rm hel}=300.2 \kms$ {\it are\/} thinner than
those at $v_{\rm hel}=357.9 \kms$, the models show much more extra-planar
emission than the data in the extreme channels.  The Power-Law model (third
column) has globally less extra-planar emission than the others (see Section
\ref{s_constraints}) but the kinematic pattern is the same.  
Thus regardless of the distribution of kick velocities, 
model extra-planar material invariably rotates too fast.
\citet{fra05} have estimated the vertical
gradient in rotation velocity in NGC\,891
to be about $15 \kms \rm kpc^{-1}$ 
(in the range $1.3< \rm z <5.2$ kpc). 
We have applied the same technique to our model cube and found a value
of $\sim 7 \kms \rm kpc^{-1}$ (a {factor of two} smaller than in the data).
{We have also inspected the behaviour of $v_{\phi}(R,z)$ in our model
and found a very similar value for the rotation velocity gradient
$\sim 6 \kms \rm kpc^{-1}$ between $R\sim 5$ kpc and $R_{\rm
cut}$ (the cut-off radius of star formation).
In the inner regions the gradient is more pronounced, beyond $R_{\rm
cut}$ the gradient become steeper close to the plane ($z\lsim 1$kpc) 
and then shallower afterwards. 
The velocity gradient inferred from the observations relates to the region
$5\kpc\lsim R\lsim R_{\rm cut}$
and has to be compared with the value of $6-7 \kms \rm kpc^{-1}$ 
derived from the model. 
In the innermost regions, the data also indicate the possibility of a
larger gradient \citep{fra05}.
}

It seems that clouds that are shot up loose part of
their angular momentum, rather than conserving it. 
A possible mechanism for this loss is interactions with 
pre-existing {\it hot\/} halo or infalling material from the IGM (see
Section \ref{s_discussion}).

\begin{figure*}
  \includegraphics[width=400pt]{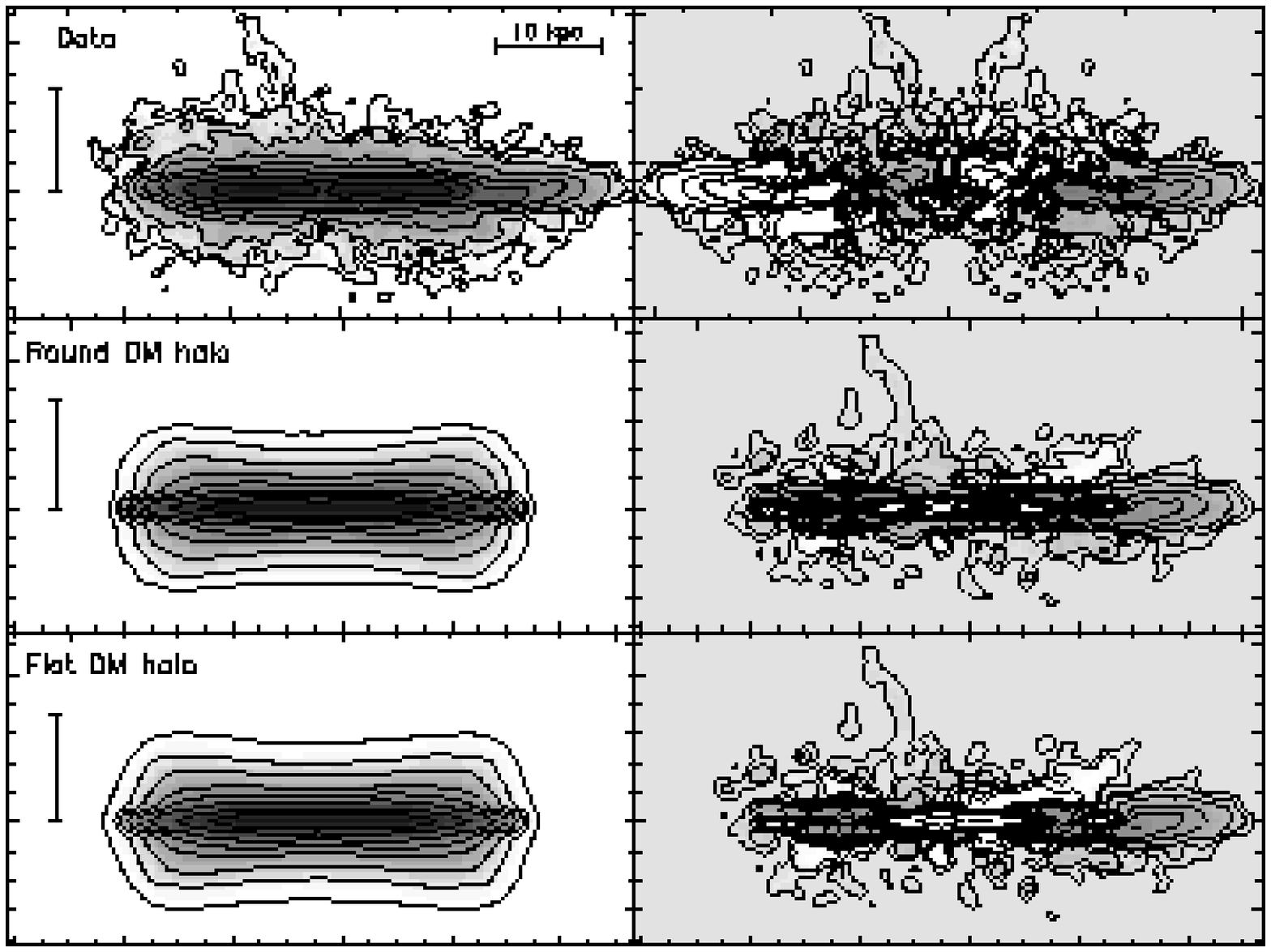}
  \caption{Comparison between the (rotated) total \hi\ map of NGC\,891
  (upper left panel) \citep{oos05} 
and those of the two dynamical models in Table
  \ref{t_effects}.
  The right panels show the residuals of the subtraction of the model
  total maps from the total \hi\ map.
  The upper right panel shows the results of subtracting a minor-axis
  mirrored \hi\ map from the original.
    In the left panels the contour levels are: 0.2, 0.45, 1, 2.5, 5, 10,
    25, $50 \mopc$. 
    {In the right panels the grey background indicates the zero level, 
darker colours imply that the measured intensity is larger than that
predicted by the model, and vice-versa for lighter colours. 
The contour spacing is the same as in the left panels.
}
\label{f_models891_tot_2}
}
\end{figure*}

\subsection{Impact  of the potential} 
\label{s_effects891}

We now investigate the effects of different galactic potentials for
NGC\,891.  The second and third rows of Fig.~\ref{f_models891_tot_2} show
the total maps obtained with our best fit models with potentials dominated
by the dark matter halo, which may be round or flat (see Table
\ref{t_massmodels891}).  

\begin{table}
 \centering
 \begin{minipage}{140mm}
   \caption{Parameters of the DM-dominated models for NGC\,891}
   \label{t_effects}
   \begin{tabular}{@{}lcc@{}}
     \hline
     Model & $h_{\rm v}$ & $M_{\rm halo}$ \\
     & $(\!\kms)$ &  ($10^9 \mo$) \\
     \hline
     Round DM halo & 70 & 1.85 \\
     Flat DM halo & 80 & 2.4  \\
     \hline
   \end{tabular}
 \end{minipage}
\end{table}

The best fit parameters of the models with maximum-DM potentials differ
slightly from those obtained with the maximum-light potential (see Table
\ref{t_effects}).  In particular, for the first model we have $h_{\rm v}=70
\kms$ and $M_{\rm halo}=1.85\times 10^9 \mo$, both lower than the values required
for the maximum-light model.  This finding  reflects the fact that in a
round potential it is easier for particles to leave the plane of the disc.
The distributions of extra-planar gas and the residuals obtained with these
two models are comparable.  They are also comparable with those obtained
with the maximum disc models.

\begin{figure}
  \includegraphics[width=240pt]{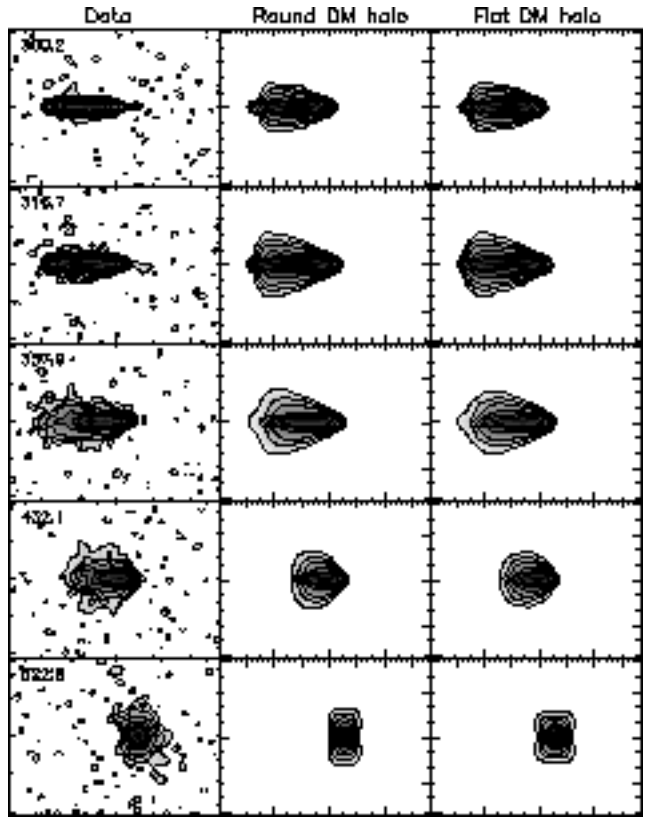}
  \caption{
    Comparison between five observed channel maps of NGC\,891 \citep{oos05}
    and those produced with the two dynamical models in Table
    \ref{t_effects}.
    The first column shows the data, heliocentric radial velocities
    are reported in the upper left corner. 
{The channel maps of the bottom row are roughly at the systemic velocity 
($V_{\rm sys}=528 \kms$).
}
    Contour levels (for data and models) are: 0.45 ($2\sigma$), 1, 2,
    5, 10, 20, 50\,mJy/beam. 
    \label{f_models891_chan_2}
}
\end{figure}

What about the kinematics?  Fig.~\ref{f_models891_chan_2} shows channel maps
equivalent to those shown in Fig.~\ref{f_models891_chan_1}.  These channel
maps differ little from those obtained with the maximum-light potential.  In
particular, the upper channels (away from systemic velocity) are again thicker
than those of the data, so the halo-dominated models fare no better in
producing the required lag in the rotation of the extra-planar gas. Thus
the dynamics of the extra-planar gas does not depend
significantly on the potential of the galaxy.

\subsection{Outflow and inflow rates} 
\label{s_inout891}

In this section we ask the question whether the model is physically feasible
in terms of mass loss and energy input.  In particular we look at the fate
of the gas pushed up from the galactic plane through the lifetime of the
galaxy.  Finally, we calculate the energy input necessary to maintain the
fountain. 

\begin{figure}
  \includegraphics[width=240pt]{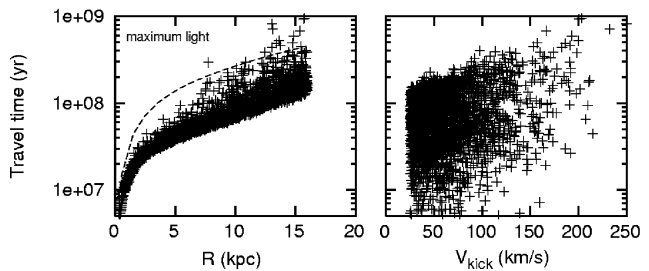}
  \caption{Travel times for particles in the maximum light potential of 
  NGC\,891 as a function of initial radius (left) and modulus of the
  kick velocity (right). 
  The dashed line in the left panel
shows the revolution time in the disc as a function
  of radius.
\label{f_times}
}
\end{figure}

The good agreement between the model and observed total-\hi\ maps guarantees
that the models have the right mass of extraplanar gas. However, the mass of
the halo gas that we obtain from eq.~\ref{eq_m_halo} is unrealistically large: for 
all models we obtain $M_{\rm halo} \simeq2 \times 10^9\mo$, which is half of
the total \hi\ mass of the galaxy and about 4 times larger than the
``observed'' halo mass.  This discrepancy arises because eq.~\ref{eq_m_halo}
counts everything that is shot up, while the observed halo mass includes
only material at distances $z$ above the plane that exceed the observational
resolution, $\sim1.3\kpc$ in the case of NGC\,891.
Indeed if we consider only the material travelling above 1.3 kpc from the 
plane we obtain a mass of $M_{\rm halo, z>1.3\kpc}\simeq6 \times 10^8 \mo$.

It is interesting at each radius to compare the rate at which material is
shot up with the rate at which it returns to the disc.
The amount of mass pushed up is related to the outflow
rate by eq.\ \ref{eq_n_halo}. 
Depending on their initial radius and kick velocity the particles
remain in the halo for a time $\tau(R,\vec{v})$.
Fig.~\ref{f_times} shows the travel times for 2000 particles 
as a function of radius (left panel) and kick velocities (right
panel) for a maximum light model.
The travel times are quite short: half to a third of
the dynamical (revolution) times at each radius (dashed line). 
They are slighly larger in a rounder (DM dominated) potential.

\begin{figure}
  \includegraphics[width=240pt]{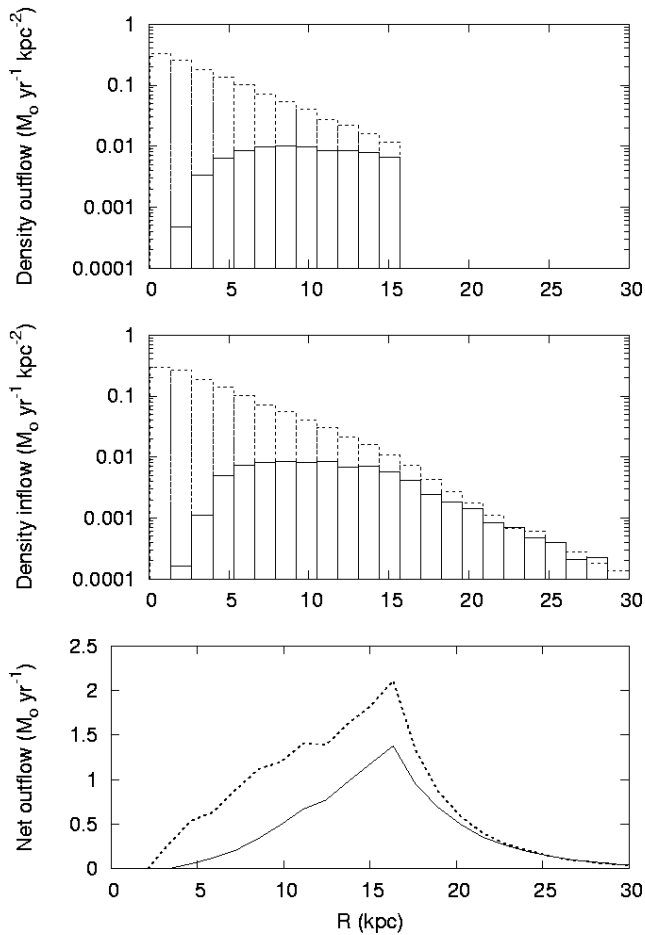}
  \caption{Rates of outflow (upper panel) and inflow (middle panel) per unit
  area of the disc of NGC\,891.  The bottom panel shows the rate at which
material flows through a cylinder of radius $R$.  Dashed lines include all
material, while full lines show only material that crosses the surfaces
$|z|=1.3\kpc$. The potential is for  maximum-light.  \label{f_inout891} }
\end{figure}

\begin{table}
 \centering
 \begin{minipage}{140mm}
   \caption{Inflow/outflow rates and energy input for NGC\,891}
   \label{t_inout891}
   \begin{tabular}{@{}lcccc@{}}
     \hline
     Model & Total outflow & Net outflow & Energy input \\
     & $(\!\moyr)$ & $(\!\moyr)$ & ($10^{39}\ergs$)\\ 
     \hline
     Maximum light & & & \\
~~~~~~~~~~~~~($z>0$) & 33.1 & 2.1 & 49.8 \\
~~~~~~~~~~~~($z>1.3$) & 5.7 & 1.4 & 19.2 \\
     Maximum DM  &  & & \\
~~~~~~~~~~~~~($z>0$) & 22.7 & 1.3 & 33.2 \\
~~~~~~~~~~~~($z>1.3$) & 5.6 & 1.0 & 15.1 \\
     \hline
   \end{tabular}
 \end{minipage}
\end{table}

Once the mass of the halo has been fixed, the orbits have been integrated
and the travel times derived, we can infer the outflow rates.
Fig.~\ref{f_inout891} shows the outflow (upper panel) and inflow (middle
panel) rates at each radius for the maximum-light model.  Dashed lines show
the flow rates at $z=0$, while solid lines show the flow of material at
$|z|=1.3\kpc$, and thus give an indication of the flow of material that would
appear as extraplanar in the observations.  Especially at small $R$ the
great majority of particles never reach $|z|=1.3\kpc$, and nearly all these
particles return to the disc within the same radial bin. Near the outside of
the star-forming disc, a significant fraction of the particles that leave
the disc pass through $|z|=1.3\kpc$.

The bottom panel of Fig.~\ref{f_inout891} shows the rate of flow of material
across a cylinder of radius $R$ as a function of $R$; the dashed curve shows
the flow for the whole cylinder and the full curve shows the flow for
$|z|>1.3\kpc$. This flow peaks at $\dot M \sim1.4\mo\yr^{-1}$ at the edge of 
the star-forming disc.  If gas is permitted to cross the plane on its first
return to the plane, the net flow is reduced to $\dot M \simeq 0.2 \mo/\yr$.
We discuss such `second-passage' models in detail in Sections \ref{s_n2403}
and \ref{s_invisibility}.  Table \ref{t_inout891} reports the global values
of outflow and inflow rates for NGC\,891, at $z=0$ and at $z=1.3\kpc$.
These values are very similar to those obtained by \citet{col02}.

From the mass outflows we can derive the total energy input required to
maintain the \hi\ halo, which turns out to be a few $\times10^{40}\ergs$.  
If we consider a supernova rate of
$\sim 0.04\yr^{-1}$ for NGC\,891 (calculated from the SFR assuming a Scalo
IMF \citep{sca86}) and an energy per supernova of $10^{51}\,$erg, the
efficiency required for transferring kinetic energy to the extra-planar gas
would be $\eta \lsim 0.039$.  
Therefore the fountain flow
required by our model is energetically plausible. 

In summary, the application of our model to NGC\,891 has led to the
following conclusions: 1) the model reproduces the vertical gas distribution
with characteristic kick velocities of about $75 \kms$; 2) the energy
requirement to maintain this mechanism is less than 4 
percent of the total energy
input from supernovae; 3) the model's main shortcoming is that it gives the
extra-planar gas too little lag; 4) this lack of lag is independent of the
choice of potential.

\section{Application to NGC\,2403} 
\label{s_n2403}

NGC\,2403 is a nearby Sc galaxy located at
a distance of $3.2\,$Mpc \citep{fre01}. 
NGC\,2403 is much less luminous ($L_{\rm B}=8.2\times10^9 \lo$)
and less massive than NGC\,891 and is in several respects similar
to M33 in the Local Group.  It has a fairly high SFR with
very bright \hii\ regions \citep{dri99} and a global SFR rate of $1.2
\moyr$ \citep{ken03}.

NGC\,2403 is inclined at $63\de$ rather than being edge-on. This intermediate
inclination has the disadvantage that we cannot directly determine the
distance from the plane at which emission occurs. It has, however, the
advantage that it enables us to distinguish radial outflow from inflow.

Recent \hi\ observations of NGC\,2403 have revealed a component of neutral
gas at lower rotation velocities ($20-50 \kms$) with respect to the rotation
of the thin disc \citep{sch00,fra01}. Three-dimensional modelling of the
\hi\ data cube has shown that this component is produced by lagging
extra-planar gas as observed in edge-on galaxies, and revealed evidence that
the extra-planar gas is flowing inward \citep{fra01}.

\subsection{Mass models for NGC\,2403} 
\label{s_massmodel2403}

We have used the \hi\ rotation curve derived by \citet{fra02}.
Since no stellar bulge is detected in this galaxy \citep{ken85},
there are just three contributors to the potential: stellar and gaseous 
discs, and DM halo.
The scale length of the stellar and gaseous discs are $2.0\kpc$
\citep{ken85} and $5.7\kpc$ \citep{fra02}. 

\begin{figure}
  \includegraphics[width=240pt]{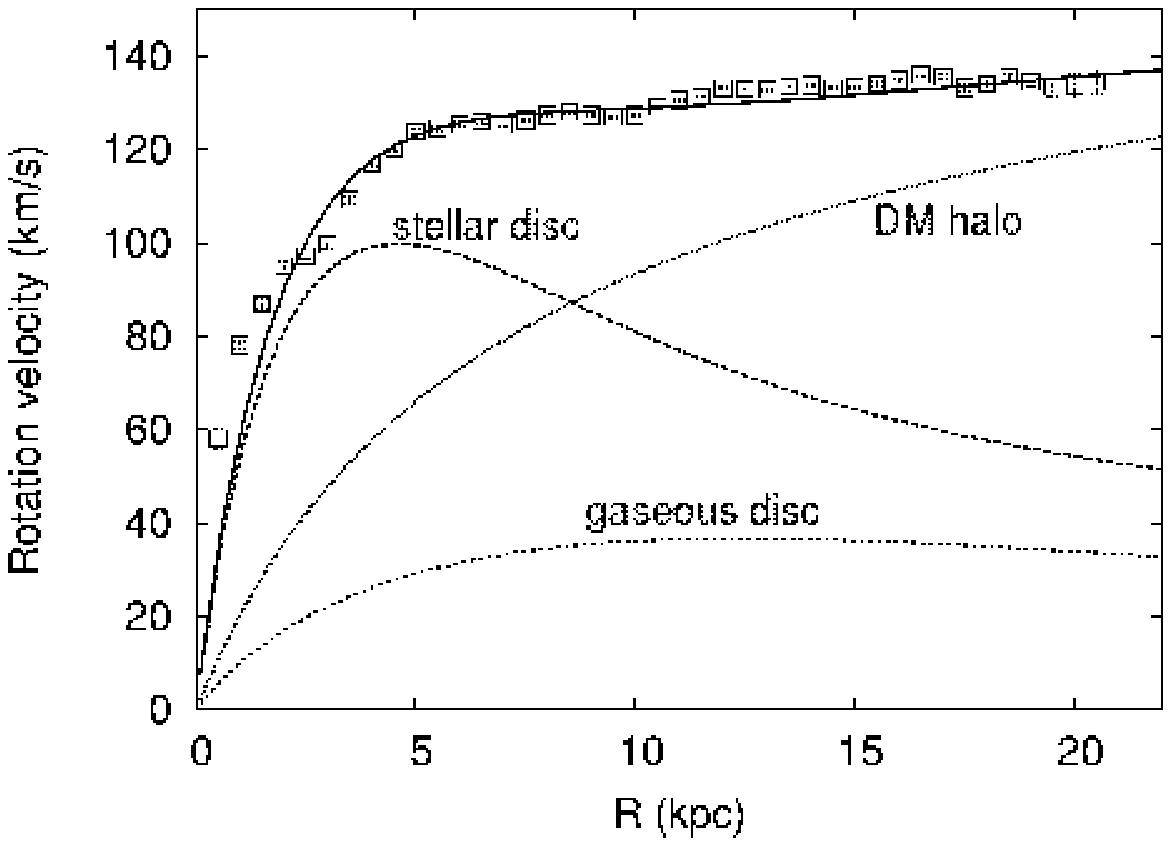}
  \includegraphics[width=240pt]{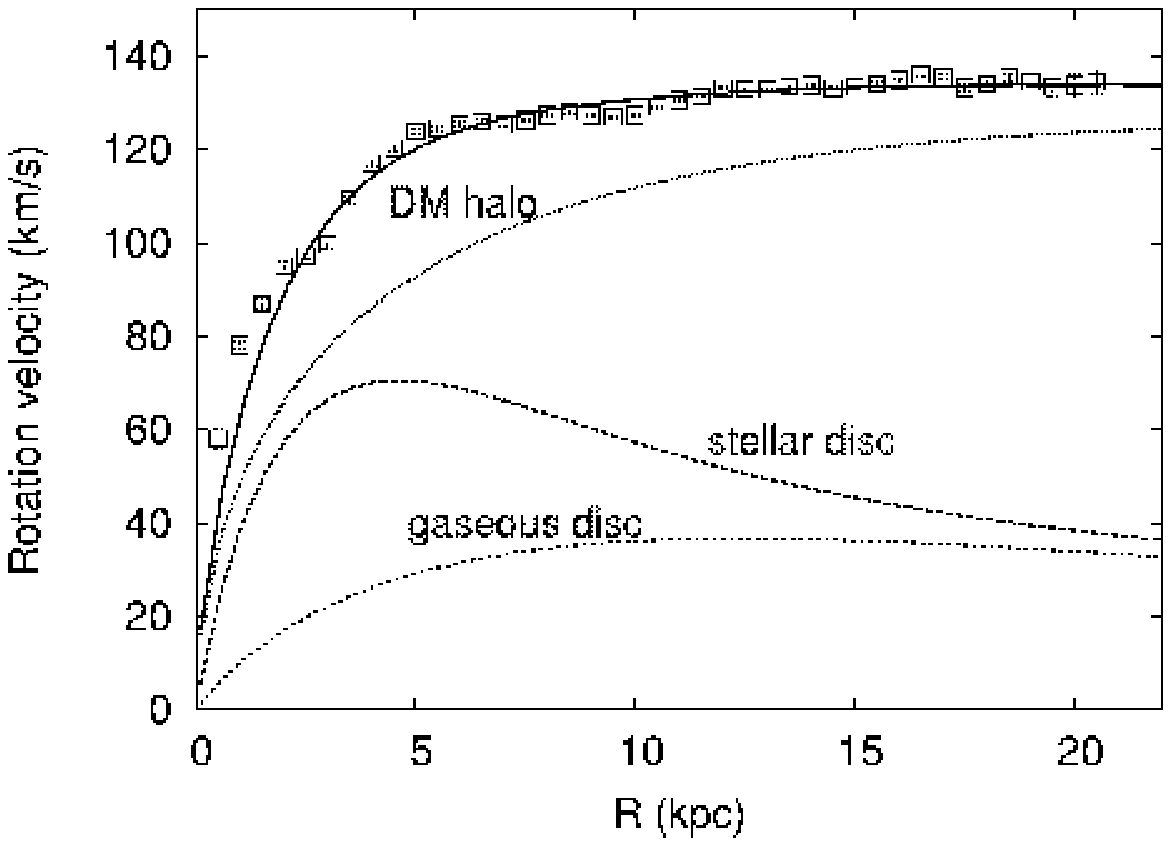}
  \caption{Fits to the \hi\ rotation curve of NGC\,2403 (squares)
  obtained with our maximum disc (top panel) and maximum halo
  (bottom panel) models. Their parameters are shown in Table 
  \ref{t_massmodels2403}.
  \label{f_massmodels2403}
}
\end{figure}

\begin{table*}
 \centering
 \begin{minipage}{140mm}
   \caption{Parameter of the mass models for NGC\,2403}
   \label{t_massmodels2403}
   \begin{tabular}{@{}lcccccccc@{}}
     \hline
     Model & $(M/L)_{\rm disc}$ & $R_{\rm d}$ & $h_{\rm d}$ &
     $\rho_{\rm 0,DM}$ & $a$ & $\gamma$ & $\beta$ & $e_{\rm DM}$\\   
     & $(\!\mo/\loB)$ & (kpc) & (kpc) & $(\!\mo\kpc^{-3}$) & (kpc) & & & \\
     \hline
     maximum light & 1.70 & 2.0 & 0.4 & 3.1$\times$10$^7$ & 4.5 & 0 &
     2 & 0 \\ 
     maximum round DM halo & 0.85 & 2.0 & 0.4 & 6.9$\times$10$^6$ &
     14.0 & 1 & 3 & 0 \\ 
     \hline
   \end{tabular}
 \end{minipage}
\end{table*}

Fig.~\ref{f_massmodels2403} (top panel) shows the fit to the rotation curve
obtained with the maximum-disc model for $M/L_{\rm B}=1.7$.  The DM halo in
the maximum-disc fit is a double power-law model (eq.\ \ref{eq_rhodpl}) with
$\gamma=0$, $\beta=2$ and $a=4.5\kpc$.  The bottom panel of
Fig.~\ref{f_massmodels2403} shows the fit to the rotation curve of NGC\,2403
obtained with our maximum-DM model, in which the stellar disc has
$M/L_{\rm B}=0.85$, half the value in the maximum-disc model.  
With this choice, the DM halo is dominant at all radii.  The other parameters
of these mass models are listed in Table \ref{t_massmodels2403}.

\subsection{Dynamics of the extra-planar gas} 
\label{s_dynamics2403}

\begin{figure}
  \includegraphics[width=240pt]{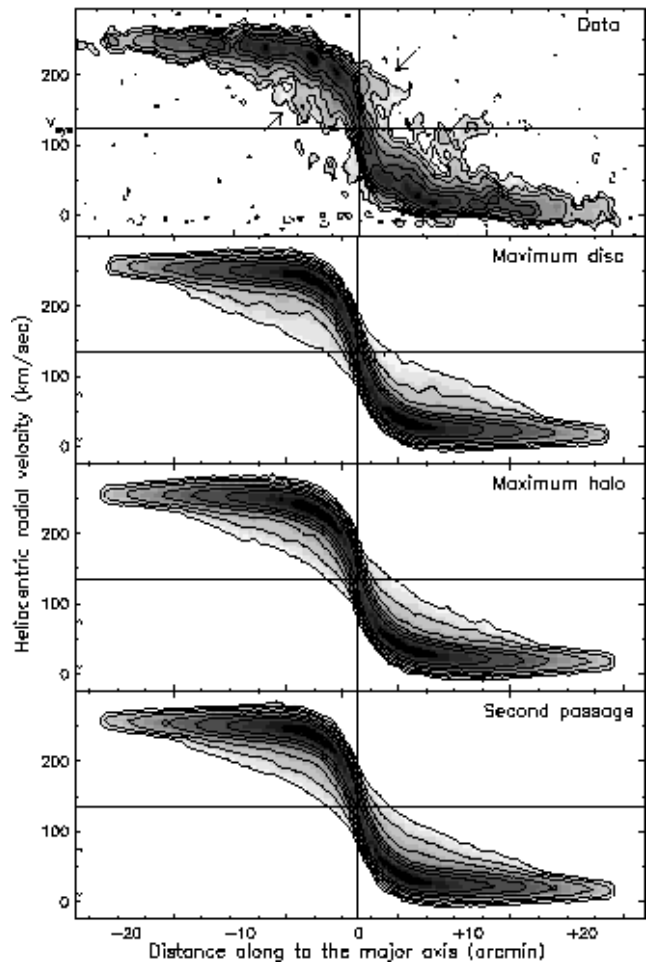}
  \caption{
    Position-velocity plot along the major axis of the spiral galaxy
    NGC\,2403 \citep{fra02} {top panel}
compared with the results of three dynamical models.
    The left arrow in the data p-v plot indicates the location of
    the extra-planar gas, the right arrow points at the {\it
    forbidden} gas (see text).
{In each panel the horizonthal lines shows the systemic velocity ($V_{\rm sys}=133 \kms$).
}
    Contour levels are: 0.5 ($\sim$2.5 $\sigma$), 1, 2, 5, 10, 20, 50 
    mJy/beam. 
    \label{f_models2403_max}
  }
\end{figure}

Both displacement of gas from the galactic plane and rotational lag shift
emission towards the systemic velocity and away from the velocities of disc
gas.  This phenomenon is apparent in a position-velocity (p-v) plot along
the major axis of the galaxy (Fig.~\ref{f_models2403_max}, upper panel).  We
use here the data of \citet{fra02} at a resolution of $30''$ ($0.46\kpc$).
The extra-planar gas in this p-v plot is visible as a tail in the region of
low rotation velocities towards the systemic velocity (indicated by the
left arrow).  It is difficult to say what percentage of this low-rotation
gas (said to form a ``beard'') is produced by lag rather than by thickness
effects \citep{sch00}.  There is also halo gas at very high velocity, as
indicated by the right arrow in Fig.~\ref{f_models2403_max}.  This is
``forbidden'' gas because it shows up in quadrants of the p-v diagram (upper
right and bottom left) that are inaccessible to gas that moves on circles in
the same sense as the disc, regardless of its distance from the plane or
speed of rotation. We defer discussion of the nature of this gas to Section
\ref{s_forbidden}.

\begin{table}
 \centering
 \begin{minipage}{140mm}
   \caption{Parameters for the models of NGC\,2403}
   \label{t_models2403}
   \begin{tabular}{@{}lcc@{}}
     \hline
     Model & $h_{\rm v}$ & $M_{\rm halo}$ \\
     & $(\!\kms)$ &  ($10^9 \mo$) \\
     \hline
     Maximum disc & 70 & 0.5 \\
     Maximum halo & 65 & 0.45  \\
     Second passage & 70 & 0.5 \\
     \hline
   \end{tabular}
 \end{minipage}
\end{table}

The lower panels of Fig.~\ref{f_models2403_max} are p-v plots derived with
three dynamical models.  All the models have a Gaussian distribution of kick
velocities with $h_{\rm v}=70$ or $65 \kms$ for maximum-disc and maximum-DM,
respectively (Table \ref{t_models2403}).  We have kept these values similar
to those found for NGC\,891 since it is reasonable to think that they are
related to physical characteristics of the ISM, rather than to the potential
of the galaxy.

In Fig.~\ref{f_models2403_max} the second panel down shows the maximum-disc
model, while the third panel down shows the maximum-DM model.  The mass
of halo gas used here is about $M_{\rm halo}\simeq5\times10^8 \mo$,
considerably lower than that needed for NGC\,891.  This halo mass is also
quite close to that derived from the data i.e.\ about $3\times10^8 \mo$.
Note, however, that the technique used to separate the halo gas from the
disc is, in this case \citep{fra02}, totally different from that used for
NGC\,891 \citep{swa97} and a direct comparison may be misleading. 

Fig.~\ref{f_models2403_max} shows that both models reproduce the global
shape of the p-v diagram along the major axis, although they cannot
reproduce the forbidden gas in the central parts of the galaxy (right arrow
in top panel); the models produce tails only in the region of
low rotation velocities, and the brightness of such tails is very similar to
that in the data.  We stress that in the case of NGC\,2403 we did not
perform any minimization of residuals but simply used values for $h_{\rm v}$
close to those obtained for NGC\,891 and tuned the halo mass and the outer
radius of star formation ($R_{\rm cut} \simeq15\kpc$) to match the p-v
diagram along the major axis.  We have experimented with different values of
these parameters and found that $h_{\rm v}$ is not tightly constrained: one
can get a good fit to the p-v plot along the major axis with values as low
as $40 \kms$.

\begin{figure*}
  \includegraphics[width=400pt]{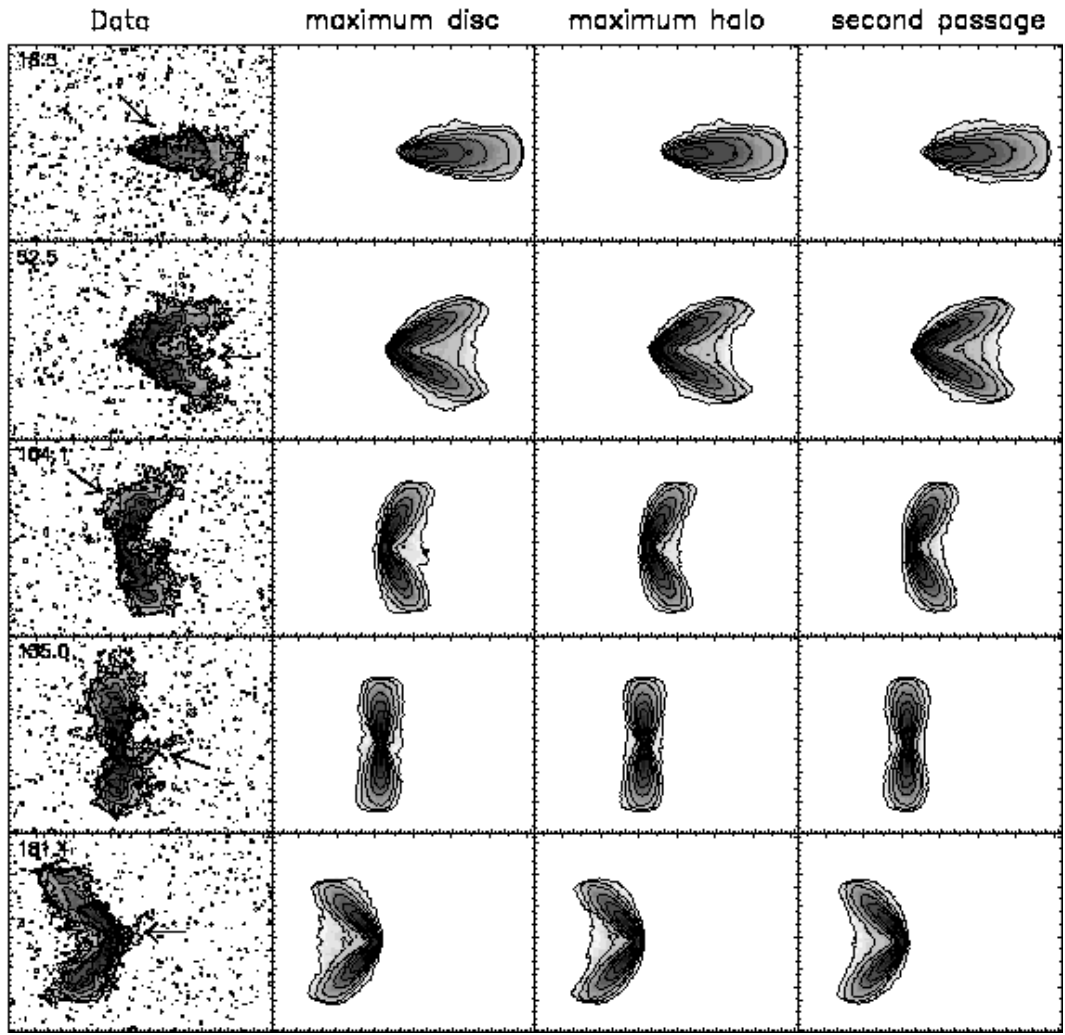}
  \caption{
    Representative channel maps for NGC\,2403 in the first column \citep{fra02} 
compared with the three 
    dynamical models in Table \ref{t_models2403}.
{Heliocentric radial velocities
    are given in the upper left corners of each row.}
    The arrows indicate various locations of the extra-planar gas (see
    text).
{The channel maps in the second row up are roughly at the systemic velocity 
($V_{\rm sys}=133 \kms$).}
    Contour levels are: 0.5 ($\sim$2.5 $\sigma$), 1, 2, 5, 10, 20, 50 
    mJy/beam. 
\label{f_models2403_chan}
}
\end{figure*}

Fig.~\ref{f_models2403_chan} compares five representative channel maps for
NGC\,2403 with those obtained with our models.  The arrows in the data
indicate features of the extra-planar gas that should be compared with the
models.  
The typical location of the extra-planar gas is that shown by the
arrow in the second row, which is for $v_{\rm hel}=52.5 \kms$, the velocity of
material that is rotating a shade more slowly than the disc gas.  
The corresponding
model maps reproduce this emission quite well.  
The arrows in the bottom two
channels show peculiar features in the extra-planar gas: in the fourth row
($v_{\rm hel}=135.0$) a massive $8\kpc$-long filament, and in the fifth row
($v_{\rm hel}=181.4$) forbidden gas (see Section \ref{s_forbidden}).
Although our smooth, axisymmetric models cannot reproduce these features, in
the third row ($v_{\rm hel}=104.1 \kms$) the maximum-disc model does
reproduce a similar feature.

The travel times for particles in the potential of NGC\,2403 are generally
longer than those found for NGC\,891 due to the lower attraction from the
disc.  As we have seen, the mass  of halo gas required to reproduce the data is
smaller.  This implies that the outflow rates are much reduced in this
galaxy with respect to NGC\,891.

\begin{table}
 \centering
 \begin{minipage}{140mm}
   \caption{Inflow/outflow rates and energy input for NGC\,2403}
   \label{t_inout2403}
   \begin{tabular}{@{}lcccc@{}}
     \hline
     Model & Total outflow & Net outflow & Energy input \\
     & $(\!\moyr)$ & $(\!\moyr)$ & ($10^{39}\ergs$)\\ 
     \hline
     Maximum disc & 3.15 & 0.23 & 3.85 \\
     Maximum halo & 2.72 & 0.21 & 2.93 \\
     Second passage & 1.70 & 0.16 & 2.05 \\
     \hline
   \end{tabular}
 \end{minipage}
\end{table}

Table \ref{t_inout2403} shows these values for the three models presented
here together with the energy input required to maintain the \hi\ halo.
These values are about an order of magnitude lower than those obtained for
NGC\,891 (Table \ref{t_inout891}).  For a supernova rate of $0.01\yr^{-1}$
\citep{mat97} and an energy per supernova of $10^{51}\,$erg, the required
efficiency is $\eta \lsim0.01$.  Hence, only a small fraction of the energy
released by supernovae is required to maintain the fountain.

\subsection{Inflow or outflow?} 
\label{s_inout2403}

\begin{figure}
  \includegraphics[width=240pt]{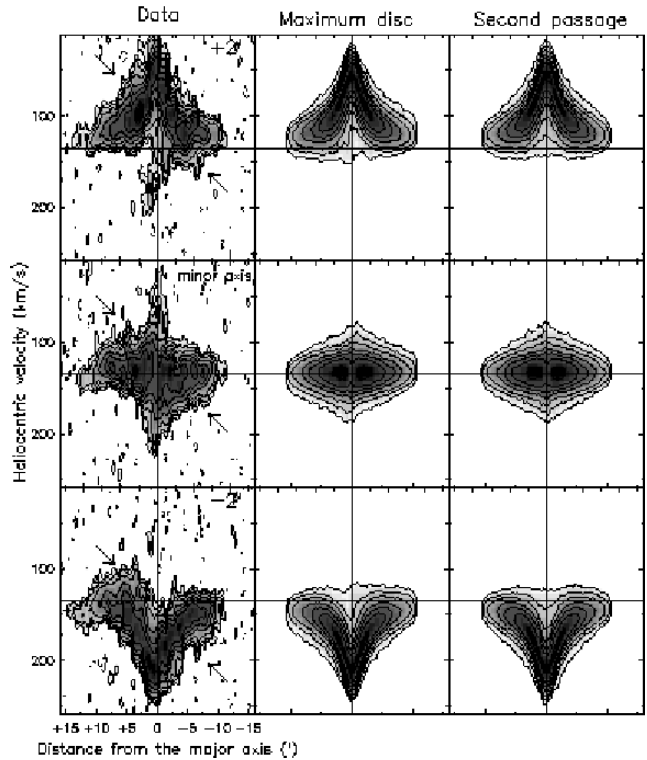}
  \caption{	
Comparison between position-velocity plots perpendicular to the major
axis of NGC\,2403 in the {first column}  \citep{fra02}
and two dynamical models.
The parameters of the model are shown in Table \ref{t_models2403}.
The arrows outline the asymmetries in the data that may be
indication of a general inflow of the extra-planar gas.
{In each panel horizonthal lines show the systemic velocity ($V_{\rm sys}=133 \kms$).
}
    Contour levels are: 0.5 ($\sim$2.5 $\sigma$), 1, 2, 5, 10, 20, 50 
    mJy/beam. 
\label{f_models2403_min}
}
\end{figure}

We now turn to the most significant discrepancy between the data and the
models of NGC\,2403.  In the third row of Fig.~\ref{f_models2403_chan} an
arrow emphasises that at velocities close to the systemic velocity ($v_{\rm
sys}=133 \kms$) the emission from extra-planar gas is rotated slightly
counter-clockwise in the channel map.  This phenomenon, can be better
appreciated in p-v plots perpendicular to the major axis.
Fig.~\ref{f_models2403_min} shows such plots for three perpendiculars: the
middle row is for the minor axis, while the plots above and below are for
$2'$ north-west and south-east of the minor axis, respectively. If the gas were
in circular rotation, these plots would be symmetric with respect to the
central vertical line (the location of the major axis).  Radial motion of
the gas inwards or outwards will break this symmetry, as is most easily
understood by considering the p-v plot along the minor axis.  Actually all
three data plots are strongly asymmetric, as indicated by the arrows.  To
determine whether the observed asymmetry corresponds to inflow or outflow,
we need to know which side of the galaxy is nearer.  Absorbing features in
the disc of NGC\,2403 and the assumption that the spiral arms are trailing
led \citet{fra01} to conclude that the south-west side (right in these
plots) is nearer. Then the asymmetry of the observed p-v diagrams implies
that the extra-planar gas is flowing in towards the galactic centre
\citep{fra02}.  Similar evidence for an inflow of the extra-planar has recently
been reported for NGC\,4559 \citep{barb05}. 

\begin{figure}
  \includegraphics[width=240pt]{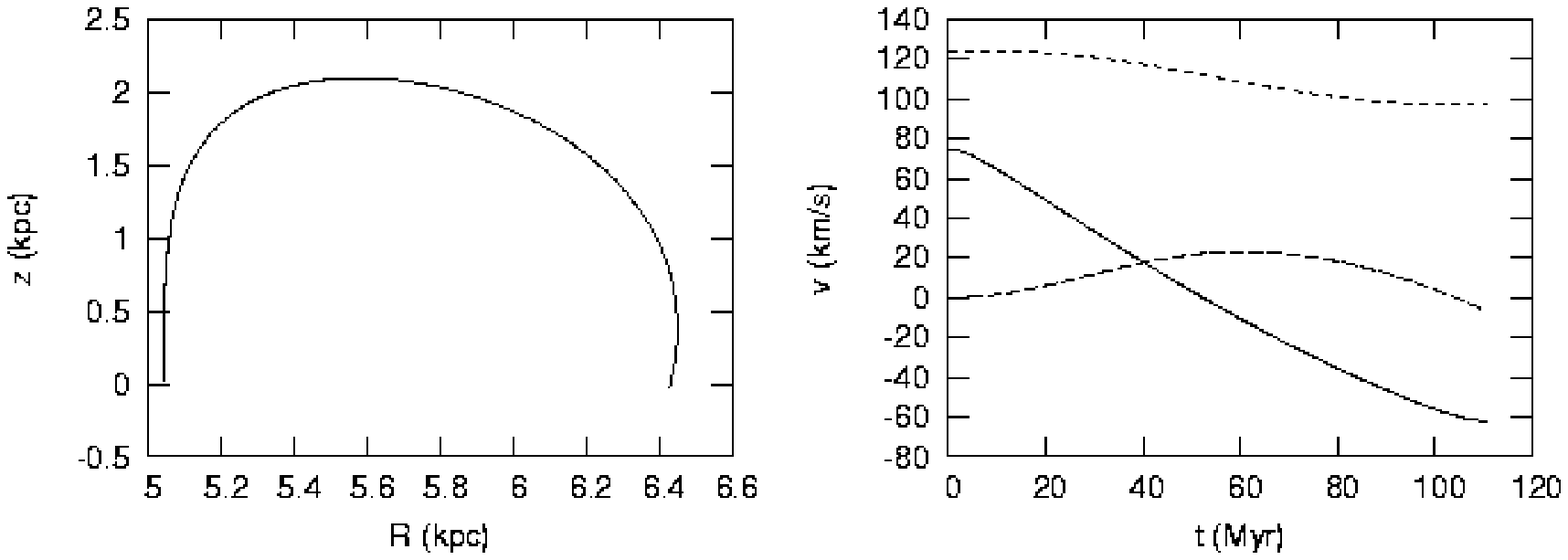}
  \includegraphics[width=240pt]{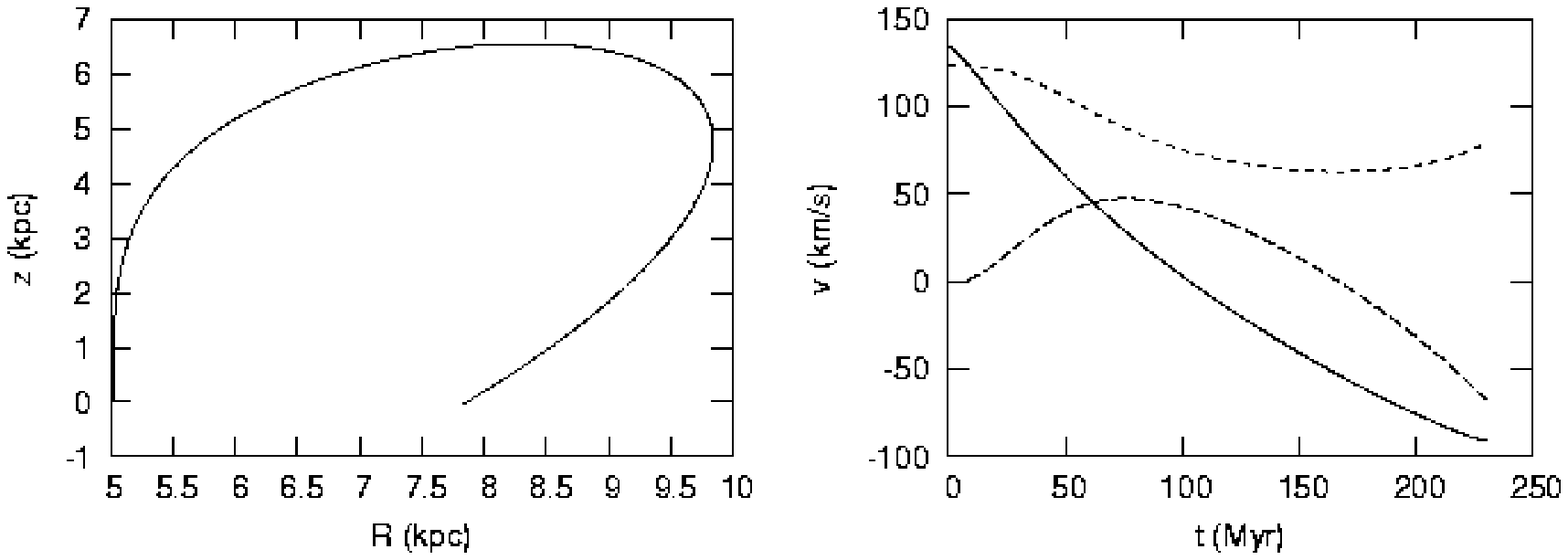}
  \includegraphics[width=240pt]{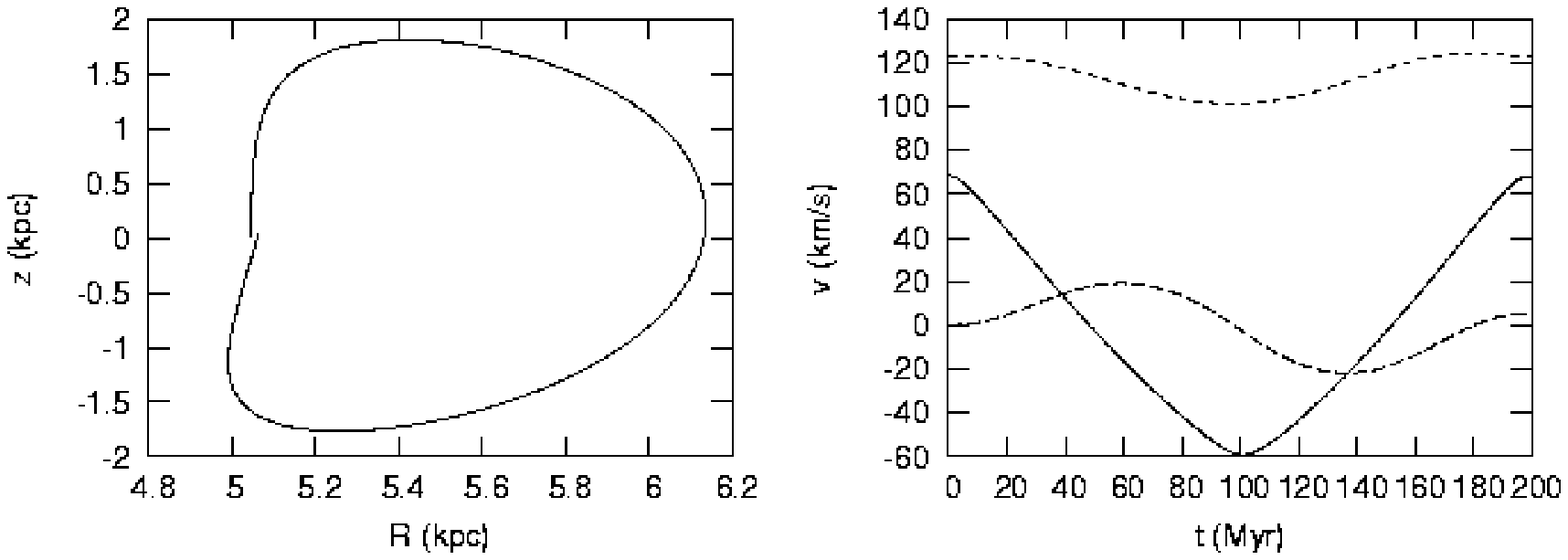}
  \caption{
    Orbits in the $(R,z)$ plane for three representative orbits in 
    the maximum-disc potential of NGC\,2403. 
    In the bottom panel the orbit passes
    through the disc at the first passage. 
Right panels: solid line=$v_{\rm z}$, long dashed line=$v_{\rm R}$, short dashed line=$v_{\phi}$.
    \label{f_orbits2403}
}
\end{figure}

The model p-v plots in the middle column of Fig.~\ref{f_models2403_min} are
slightly asymmetric in the opposite sense to the data plots. This phenomenon
reflects the fact that in the models there is a net {\it outflow\/} of halo
gas.  It is easy to understand why the fountain model predicts an outflow: clouds
that form the halo are shot upwards nearly perpendicular to the disc from
platforms that are moving close to the local circular speed. Consequently, the
clouds start from pericentre.  They return to the disc after half a period
of vertical oscillation, which is comparable to half a period of radial
oscillation.  Hence clouds are near apocentre when they return to the disc,
and have spent most of the intervening period moving outwards.
Fig.~\ref{f_orbits2403} illustrates this reasoning by showing the $(R,z)$
coordinates along three representative orbits in the maximum-disc potential
of NGC\,2403.  Each trajectory starts at $R\simeq5\kpc$, and from top to
bottom the initial vertical velocity is $v_{\rm z}\simeq70$, 140 and $70
\kms$. It is striking that the outflow inherent in the fountain model
produces an asymmetry in the p-v diagrams of Fig.~\ref {f_models2403_min}
that is appreciably smaller than the opposite asymmetry apparent in the
data. Thus the data imply really significant inflow.

Is it possible to reconcile the fountain model with the observed inflow of
halo gas?

We have seen that ejected gas is bound to flow outwards for most of the time
prior to its return to the disc plane. If the gas disc has a sufficiently
low surface density at the (large) radius of this impact, a cloud may pass
clean through the disc. In this case it will contribute to halo-gas emission
with $\dot R<0$ for most of the time until it hits the high-surface-density
inner gas disc. The panels labelled `second passage' in
Figs~\ref{f_models891_tot_2} and \ref{s_dynamics2403} to
\ref{f_models2403_min} show what the data would look like in
this case. In particular, the column on the extreme right of
Fig.~\ref{f_models2403_min} shows that permitting clouds to pass through the
outer disc restores symmetry to the p-v diagrams for perpendiculars to the
major axis of NGC\,2403, rather than introducing the required sense of
asymmetry. This makes perfect sense physically, and suggests that our only
hope of accounting for the observed inflow with the fountain model involves
both allowing clouds to pass through the disc, and arguing that they are
visible only during the second halves of their orbits.  This condition would
be satisfied if outflowing clouds were highly ionized (perhaps because they
are hot), and the clouds become predominantly neutral near apocentre,
perhaps as a result of radiative cooling (Section~\ref{s_invisibility}).

\subsection{Halo substructures and forbidden gas} 
\label{s_forbidden}

In previous sections we have compared the data of NGC\,2403 with smooth
axisymmetric pseudo-data cubes produced with our models.  Although there is
a general agreement between the two, Fig.~\ref{f_models2403_chan}
demonstrates that the data contain features that are not reproduced by the
models.  In particular several gas complexes are detected as discrete
features or substructures with masses from about $10^5 \mo$ (limit of the
data) to $10^7 \mo$ (filament visible in channels at $v_{\rm hel}=104.1$
and $v_{\rm hel}=135.0 \kms$). Moreover some of these extra-planar gas
structures are found at anomalous velocities (forbidden gas, see the map
$v_{\rm hel}=181.4 \kms$ and Fig.~\ref{f_models2403_max}).  Can these
features be produced in a galactic fountain scheme or are they evidence of
gas accretion?

We can think of two ways to obtain these features in our models.  The first
is to increase the kick velocities in the centre of the galaxy.  This will
produce gas with large vertical velocities that may end up in the forbidden
quadrants of the p-v plot along the major axis.  The second is to increase
substantially the number of particles pushed up in the central regions by
increasing the parameters $\alpha_{\rm SF}$ in eq. \ref{eq_schmidt}.  Indeed
by inspecting the model cubes we found that at low density levels some gas
is already present at forbidden velocities but below the threshold of the
lowest contour.  Both proposals have reasonable physical explanations.
First, if many supernovae go off in a limited spatial region (a bright
\hii\ region), then the supernovae may power a supershell at higher
expansion velocities.  The second idea is also plausible considering the
uncertainties in the Schmidt law and the fact that star formation may well
depend on several other parameters in addition to gas density. 

\begin{figure}
  \includegraphics[width=240pt]{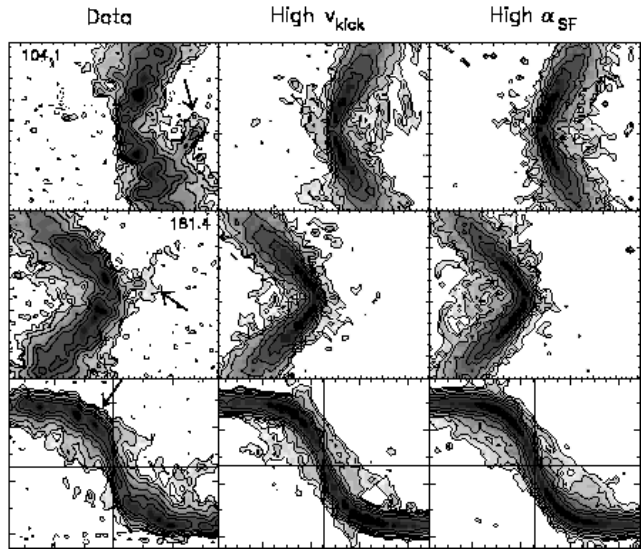}
  \caption{
    Comparison between two channel maps and a p-v plot along the major
    axis for NGC\,2403 {(left column)} and two models. 
    The two models are obtained by increasing the kick velocity in the
    centre of the galaxy (second column) and increasing the density of
    fountain particles in the centre (third column). 
    The arrow indicate interesting features in the data (see text). 
    Contour levels are: 0.5 ($\sim$2.5 $\sigma$), 1, 2, 5, 10, 20, 50 
    mJy/beam. 
\label{f_models2403_chan_2}
}
\end{figure}

The left column of Fig.~\ref{f_models2403_chan_2} shows, from top to bottom,
two observed channel maps and the major-axis p-v plot. The other two columns
show the corresponding predictions for two models. In the case of the centre
column, the ejection velocity is $h_{\rm v}=120 \kms$ at $R<4\kpc$, comparable
to the circular speed, $v_{\rm circ}=130 \kms$. The right-hand column shows
the result of assuming a high SFR, $\alpha_{\rm SF}=3$, at $R<4\kpc$.  In
both cases we have enhanced substructure by reducing the number of clouds
employed in the simulations.  The top row of panels shows that both models
produce features at anomalous velocities that are not unlike some of those
observed.  The second row of panels shows that the models can even produce
emission at forbidden velocities, although not to the extent observed (arrow
in left-hand panel). Unfortunately, the major-axis p-v diagrams of the
bottom row reveals a serious problem with the models: the data show very
little gas at speeds larger than circular (arrow at left) whereas the models
do show such emission, especially in the case of large $h_{\rm v}$ (middle
column).

We summary that there is little promise of accounting for very anomalous 
emission (forbidden gas)
within the context of the fountain model. 

In conclusion, we have shown that 1) the basic distribution and
kinematics of extra-planar gas in NGC\,2403 can be reproduced with the
same model used for NGC\,891 and the same characteristic kick
velocity ($\sim70 \kms$); 2) the energy input required by this model
is less than 1 percent the input from supernovae; 3) the models fail to
reproduce the global inflow of extra-planar gas seen in the data, 
predicting instead an outflow (or no flow); 4) high velocity and
massive substructures are not easily produced by the fountain model.

\section{Phase-change models}
\label{s_invisibility}

\begin{figure*}
  \includegraphics[width=400pt]{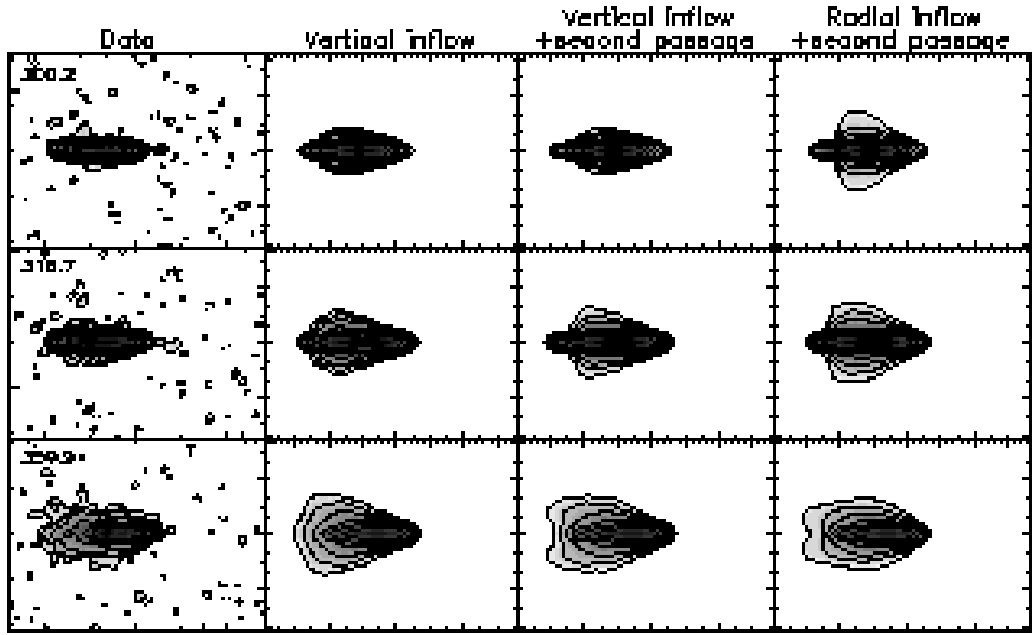}
  \caption{
    Comparison between three channel maps for NGC\,891 in the left column \citep{oos05}
and three phase-change models.
    The first column shows the data, heliocentric radial velocities
    are reported in the upper left corner {($V_{\rm sys}=528 \kms$)}.
    Contour levels (for data and models) are: 0.45 ($2\sigma$), 1, 2,
    5, 10, 20, 50\,mJy/beam. 
\label{f_models891_chan_4}
}
\end{figure*}

The models presented so far have two major failures in reproducing the
dynamics of the extra-planar gas.  They do not produce enough lag (for
NGC\,891) and they produce a general radial outflow of the extra-planar gas
contrary to what is observed in NGC\,2403.  If these failures cannot be
solved in a galactic fountain scheme, mechanisms such as accretion of
material from the IGM, will be required to explain the kinematics of the
extra-planar gas.  Here we explore the only remaining possibility to
reconcile the fountain models with the data.  This is that clouds are
visible only during the second (infalling) parts of their orbits, because
they leave the plane highly ionized and only later become predominantly
neutral. 

We have considered four models of this kind.  In the first class of models
(vertical infall) the particles are {\it switched on} (become visible as
\hi\ clouds) when $\dot z=0$ and they attain maximum height.  In this
scheme we considered two subclasses, with particles stopped at the first or
the second passage through the disc.  The other class of models (radial
infall) is constructed by switching on the particles at apocentre ($\dot
R=0$).  This conjecture ensures the maximum contribution possible from
radially infalling material.  However, only particles with high initial kick
velocities reach their apocentre before the first passage through the disc
(see Figure \ref{f_orbits2403}).  Therefore the only radial-infall model
that produces reasonable results is the one with particles stopped at the
second passage.

\begin{figure*}
  \includegraphics[width=400pt]{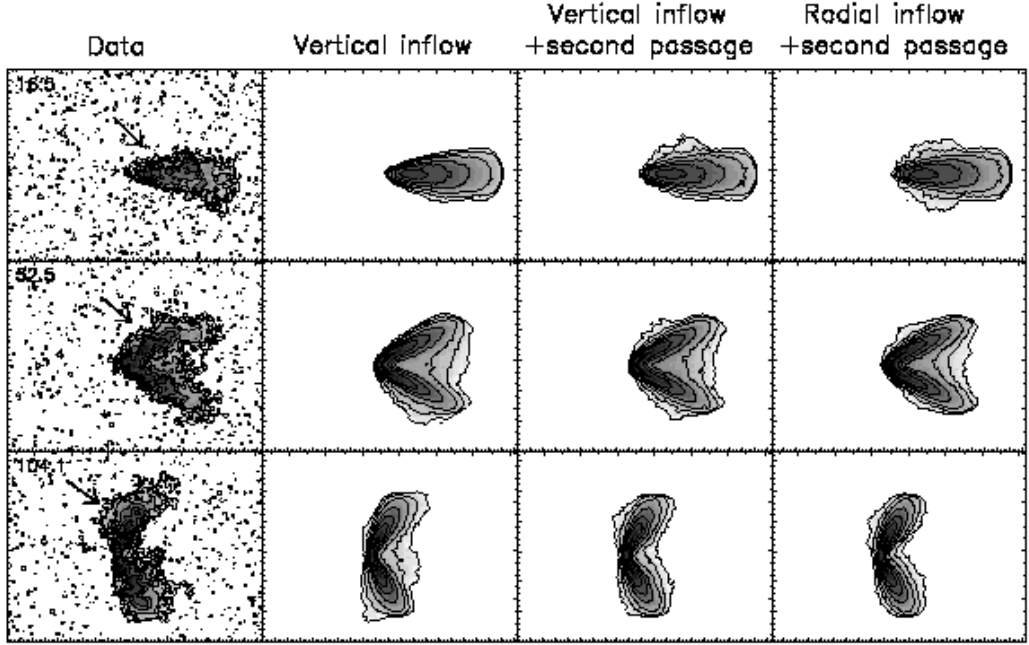}
  \caption{
    Comparison between three channel maps for NGC\,2403 (left column)
and three phase-change models.
{The first column shows the data, with heliocentric radial velocities
    in the upper left corner ($V_{\rm sys}=133 \kms$)}.
    Contour levels are: 0.5 ($\sim$2.5 $\sigma$), 1, 2, 5, 10, 20, 50 
    mJy/beam. 
\label{f_models2403_chan_4}
}
\end{figure*}

In Fig.~\ref{f_models891_chan_4} we show the comparison between three 
channel maps for NGC\,891 and those obtained with
the three phase-change models described above.
The parameters of these models were derived by minimizing the residuals
in the total \hi\ map and they are reported in Table \ref{t_invisible891}.
The cut off radius for star formation is $R_{\rm cut}=13.5\kpc$ for all
these models.
All the models have  maximum-light potentials.

\begin{table}
 \centering
 \begin{minipage}{140mm}
   \caption{Parameters of the phase-change models for NGC\,891}
   \label{t_invisible891}
   \begin{tabular}{@{}lcc@{}}
     \hline
     Model & $h_{\rm v}$ & $M_{\rm halo}$ \\
     & $(\!\kms)$ &  ($10^9 \mo$) \\
     \hline
     Vertical inflow & 75 & 2.7 \\
     Vertical inflow + second passage & 80 & 2.5  \\
     Radial inflow + second passage & 90 & 2.4 \\
     \hline
   \end{tabular}
 \end{minipage}
\end{table}

From Fig.~\ref{f_models891_chan_4} it is clear that the radial-inflow model
(rightmost column) is immediately ruled out because it produces fast
rotating extra-planar gas.  This is due to the fact that beyond the
apocentre the azimuthal component of the velocity rises again (see bottom
panel of Fig.~\ref{f_orbits2403}).  The two vertical-inflow models are
more acceptable.  They produce slighly more lagging material than the models
without a phase change (Figs~\ref{f_models891_chan_1} and
\ref{f_models891_chan_2}).  This arises because we have removed the
contribution from the first (outflowing) part of the orbits, on which the
lag is very limited (Fig.~\ref{f_orbits2403}).

\begin{table}
 \centering
 \begin{minipage}{140mm}
   \caption{Parameters of the phase-change models for NGC\,2403}
   \label{t_invisible2403}
   \begin{tabular}{@{}lcc@{}}
     \hline
     Model & $h_{\rm v}$ & $M_{\rm halo}$ \\
     & $(\!\kms)$ &  ($10^9 \mo$) \\
     \hline
     Vertical inflow & 70 & 0.5 \\
     Vertical inflow + second passage & 70 & 0.5  \\
     Radial inflow + second passage & 80 & 0.6 \\
     \hline
   \end{tabular}
 \end{minipage}
\end{table}

Consider now NGC\,2403.  Fig.~\ref{f_models2403_chan_4} shows the comparison
between three channel maps and the same three models used for NGC\,891.  The
parameters of these models are shown in Table \ref{t_invisible2403}.  The
cut-off radius is $R_{\rm cut}=11\kpc$, and all models use the maximum-disc
potential.

\begin{figure}
  \includegraphics[width=240pt]{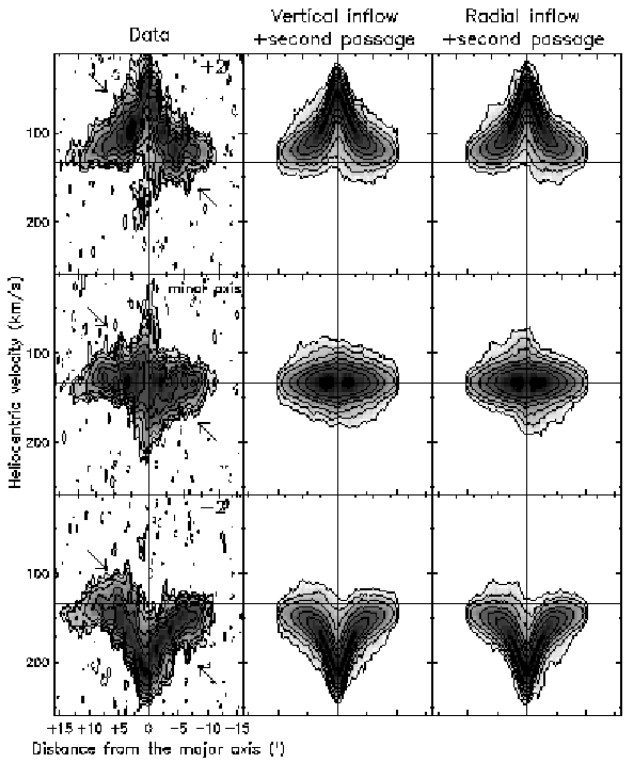}
  \caption{
    Comparison between p-v plots parallel to the minor axis of NGC\,2403 
{(left column)} and those obtained with two phase-change models.
The arrows outline the asymmetries in the data that may be
indication of a general inflow of the extra-planar gas.
    Contour levels are: 0.5 ($\sim$2.5 $\sigma$), 1, 2, 5, 10, 20, 50 
    mJy/beam. 
\label{f_models2403_min_4}
}
\end{figure}

The arrows in Fig.~\ref{f_models2403_chan_4} indicate peculiar locations
for extra-planar gas in the data.  In particular the arrow in the bottom
channel map ($104.1 \kms$) indicates a feature produced by radial inflow
motions.  Clearly only the models with particles stopped at the second
passage produce a similar pattern, whilst the vertical-inflow model displays
the opposite pattern.  The same effect is indicated by the arrow in the
channel map in the middle row ($52.5 \kms$) and the second-passage models
seem to show the correct pattern.  However, these second-passage models are
problematic in the top channel map ($16.5 \kms$) in that they show extended
tails of emission not seen in the data.  This emission is produce by
vertical motions (inflow and outflow) combined with thickness and lag.  This
seems to indicate that, although the data require a large amount of inflow,
the dominant component of this inflow should be radial rather than
vertical.

Finally, consider Fig.~\ref{f_models2403_min_4}.  This is the analogue of
Fig.~\ref{f_models2403_min} for the two second-passage models described
here.  The asymmetric pattern shown in the data (evidence of radial inflow)
is now visible in the model p-v plots.  In particular in the radial-inflow
model which is indeed the most extreme inflow model that we can produce.
However even in this extreme (probably unrealistic) model, there is less
asymmetry than in the data.  This shows that the
amount of inflow required by the data is very large.

In conclusion, combining the results for NGC\,891 and NGC\,2403, the only
promising model is the one with vertical-inflow and a second passage because
the radial-inflow model produces fast rotating extra-planar gas and the
model without second passage shows outflow instead of inflow.  However, even
this vertical-inflow model does not produce as much lag (for NGC\,891) and
inflow (NGC\,2403) as the data require.  Hence, we conclude that selecting
portions of the particle orbits does not make it possible to reproduce the
data with the fountain model.  There is an intrinsic need for low angular
momentum and radially infalling material that is not present in a pure
galactic-fountain scheme.

\section{Comparison with previous work} 

{ \cite{cor88,cha89} suggested that the extended \hi\ discs of galaxies might
be built up from gas that is shot out from interior, star-forming regions
and later settles to circular orbits of large radius. This proposal requires
some ejected gas to acquire significant extra angular momentum per unit mass
about the galaxy's symmetry axis, $J_z$, as it is expelled from the
star-forming region. We have found that the observations require the outflow
to be narrowly collimated upwards, and in this case none of the ejected gas
will increase its $J_z$ to the large values required for formation of an
extended \hi\ disc. Hence our work effectively rules out this proposal.

Our phase-change models are essentially variants of the classic
galactic-fountain model of \citet{bre80}.  In that model, star-formation in
the disc drives a hot wind that cools as it rises, with the result that
clouds of \hi\ eventually form in it, and fall back to the disc. The main
difference between our model and Bregman's is that in the latter a pressure
gradient in the coronal gas steadily accelerates the gas over a significant
distance, while in our calculations the gas is abruptly accelerated within
the disc and subsequently coasts under gravity.  Crucially, in both models
the specific angular momentum $J_z$ of gas is conserved in the flow because
the dynamics is axisymmetric. Consequently, in both models the rotation
speed of gas at cylindrical radius $R$ will be $v_c(R_0)R_0/R$, where $R_0$
is the radius from which the gas was shot upwards. It follows that in
Bregman's model, as in ours, the rotational lag of the halo gas is likely to
to be too small.

Our work closely resembles that of \citet{col02}, who also considered the
dynamics of ballistic gas clouds.  A difference in the coding is that we
integrate orbits in the $(R,z)$ plane and then uniformly distribute them in
azimuth.  This allows us to have a larger number of ``particles'' with a
limited number of integrations, drastically reducing discreteness noise.
However, the main difference between our work and that of \citet{col02} is
in the comparison with the data.  Collins et al.\ extracted average
projected velocities from the models, and compared them with the mean
velocities of ionized gas derived from \ha\ long-slit observations of
NGC\,5775 and NGC\,891.  The problem with this approach is that it does not
exploit the wealth of information that is contained in the line-of-sight
velocity distribution.  The work of Collins et al.\ has been recently
revised by \citet{hea05}, who applied the same ballistic model of
\citet{col02} to Fabry-Perot observations of NGC\,5775.  They found that the
vertical gradient in rotation velocity predicted by the model is shallower
than that measured from the data, in agreement with our results.

Finally, we stress that the distinctive feature of the work presented here
is that we attempt to reproduce directly the whole \hi\ data-cube of a
galaxy.  Our models are constructed to match the raw data without performing
any data analysis, and this procedure assures a full control of projection
and resolution effects.  }

\section{Concluding remarks} 
\label{s_discussion}

We have presented a model for the dynamics of the neutral gas outside
the plane of spiral galaxies (extra-planar gas).
We have explored the role of internal (stellar activity) processes in
the formation and maintenance of the gaseous halo. 
In our model, we have considered orbits of particles expelled from the 
disc by supernova explosions and have integrated their paths through
the halo.
We have generated output pseudo-data cubes and compared them with the
data of two galaxies with a large amount of extra-planar gas (NGC\,891
and NGC\,2403). 

The main results are:
\begin{itemize}
\item{the vertical gas distribution in NGC\,891 is well reproduced by
the models with characteristic kick velocities $v_{\rm
kick}\simeq70-80 \kms$;
}
\item{the same characteristic kick velocity also reproduces accurately
the distribution and lag of the extra-planar gas in NGC\,2403 (as seen 
in the p-v plot along the major axis);
}
{
\item{we were able to constrain the opening half-angle of the upward flow
for a typical superbubble chimney to be $\theta \lsim 15 \de$; }
}
\item{the energy input required to maintain the gaseous haloes in the
two galaxies is $<4$ percent of the energy released by supernovae;
}
\item{the actual shape of the galactic potential plays a minimal
role in the dynamics of the extra-planar gas;
}
\item{the main problem with our model it that it produces only half
the required lag  of the extra-planar gas in NGC\,891,
so there is a need for low angular momentum material (loss of angular
momentum) to reconcile the models with the data;
}
\item{a second problem is that our model predicts an outflow of the
extra-planar gas whilst,
in NGC\,2403, the data indicate an inflow;
}
\item{high velocity massive substructures, observed in NGC\,2403 and
in NGC\,891, are difficult to obtain in a fountain model and may
be the signature of accretion from IGM. 
}
\item{allowing the clouds to be invisible (ionised) for the outflowing
part of their orbits does not solve the main problems of the models.
}
\end{itemize}

We have seen how our fountain model is able to reproduce accurately
some features of the extra-planar gas (such as its vertical
distribution) and it is energetically consistent. 
However it has two main failures that clearly indicate that a
(collision-less) fountain alone cannot fully account for the dynamics
of the extra-planar gas.
The need for low angular momentum material seems to suggest that one
has to take into account interactions between the fountain flow and a
hot gaseous halo and/or accretion material.
The latter possibility seems particularly attractive. 
The accretion of material with low angular momentum (zero on average)
that interacts with the fountain gas may cause the loss of
angular momentum necessary to reproduce the lag observed in NGC\,891. 
Moreover it may add the  component of inflow velocity to the
extra-planar gas needed to reconcile the predictions of the models
with the observed inflow in NGC\,2403.
{This possibility will be explored in a forthcoming paper.}

Finally, another topic to pursue is that of our Galaxy. 
The features detected in NGC\,891 and other external galaxies are very
likely to be the analogous of the Intermediate and High Velocity
Clouds (IVCs and HVCs) of the Milky Way.
In the Milky Way we have very detailed information about the
kinematics of these features, in particular their inflow motion
\citep{wak97}, 
but also about their metallicity that,
in the case of Complex C, has been found to be of $\sim0.2$ solar
\citep{wak99,tri03}.
Such a metallicity, intermediate between a primordial value and a
galactic (roughly solar) value, may indicate a mixing process between
disc (fountain) gas and the accretion material.
The study the Milky Way halo will be a further application of
our model in the near future. 
%(Fraternali \& Binney, in preparation, Paper III).

\section*{Acknowledgments}

F.F. gratefully acknowledges support from Marie Curie
Fellowship (MEIF-CT-2003-501221).
We thank Renzo Sancisi for helpful comments and advice.
The WSRT is operated by the Netherlands Foundation for Research in
Astronomy (ASTRON) with the support from the Netherlands Foundation for
Scientific Research (NWO).

{}

\bsp

\label{lastpage}

\end{document}